\documentclass[submission, Phys]{SciPost}

\hypersetup{
    colorlinks,
    linkcolor={red!50!black},
    citecolor={blue!50!black},
    urlcolor={blue!80!black}
}

\usepackage[utf8]{inputenc}
\usepackage[T1]{fontenc}
\usepackage[english]{babel}
\usepackage{amsmath,amssymb,amsthm}
\usepackage{mathrsfs,dsfont}
\usepackage{mathdesign,bm}
\usepackage{mathtools, ifthen, xparse}
\usepackage[shortlabels]{enumitem}
\usepackage{xcolor}
\usepackage{wrapfig}
\usepackage{csquotes}
\MakeOuterQuote"
\usepackage{url}
\usepackage[capitalize]{cleveref}
\usepackage{caption,subcaption}
\usepackage{tikz}
\usepackage{pgfplots}
\usepackage{authblk}
\usepackage{fancyhdr} 
\fancypagestyle{plain}{
    \fancyhf{} 
    \fancyfoot[C]{\thepage}

}
\def\lie{{\mathfrak{g}}}
\def\diff{\mathrm{d}}
\def\Length#1{{\operatorname{L}[#1]}}
\def\Ad{\operatorname{Ad}}
\def\dU{A}
\def\Nelson{{\Delta}}
\def\Sp{\operatorname{Sp}}
\def\Spec{\operatorname{Spec}}
\renewcommand\dagger{*}
\def\rmi{\mathrm{i}}
\def\rme{\mathrm{e}}
\def\numop{{N}}
\def\states{{\mathfrak{S}}}

\newcommand{\traceclass}[1][\mathcal{H}]{\mathcal{T}\rb*{#1}}

\newcommand{\covering}[1][{G}]{{\tilde{#1}}}
\newcommand{\centext}[1][G]{\mathcal{#1}}

\newtheorem{thm}{Theorem}
\newtheorem*{thm*}{Theorem}
\newtheorem{lem}[thm]{Lemma}

\newtheorem{prop}[thm]{Proposition}

\newtheorem*{prob*}{Problem}
\newtheorem{defin}[thm]{Definition}

\theoremstyle{definition}
\newtheorem{rem}[thm]{Remark}

\theoremstyle{remark}
\newtheorem*{ex*}{Example}
\newtheorem*{rem*}{Remark}

\newcommand*{\1}{\text{\usefont{U}{bbold}{m}{n}1}}

\renewcommand{\bar}[1]{\overline{#1}}
\newcommand\eps\varepsilon

\renewcommand\H{\mathcal H}
\newcommand\K{\mathcal K}
\newcommand\NN{\mathbb N}
\newcommand\placeholder{\,\cdot\,}
\newcommand\RR{\mathbb R}

\DeclareMathOperator\tr{tr}

\newcommand\oo\infty
\newcommand\restrictedto\upharpoonright
\newcommand\ZZ{\mathbb Z}

\NewDocumentCommand\TC{o}{{\IfNoValueTF{#1}{\mathcal{T}(\mathcal{H})}{\mathcal{T}(#1)}}}
\newcommand\boundedoperatorsymbol{{\mathcal B}}
\NewDocumentCommand\BO{mo}{ \IfNoValueTF{#2}{\boundedoperatorsymbol(#1)}{\boundedoperatorsymbol(#1,#2)}}

\def\up#1{^{(#1)}}

\newcommand\mat[1]{ \begin{pmatrix} #1 \end{pmatrix} }

\renewcommand\limsup\varlimsup
\renewcommand\liminf\varliminf
\DeclareMathOperator\id{id}

\DeclareMathOperator\SO{SO}
\let\O\relax
\DeclareMathOperator\O{O}
\DeclareMathOperator\SU{SU}

\newcommand\dom{\operatorname{dom}}

\let\lim\relax
\NewDocumentCommand\lim{o}{\IfNoValueTF{#1}{\mathop{\textup {lim}}}{\mathop{\textup{{$ #1 $}-lim}}}}

\DeclareFontFamily{U}{matha}{\hyphenchar\font45}
\DeclareFontShape{U}{matha}{m}{n}{ <-6> matha5 <6-7> matha6 <7-8> matha7 <8-9> matha8 <9-10> matha9 <10-12> matha10 <12-> matha12 }{}
\DeclareSymbolFont{matha}{U}{matha}{m}{n}
\DeclareFontFamily{U}{mathx}{\hyphenchar\font45}
\DeclareFontShape{U}{mathx}{m}{n}{ <-6> mathx5 <6-7> mathx6 <7-8> mathx7 <8-9> mathx8 <9-10> mathx9 <10-12> mathx10 <12-> mathx12 }{}
\DeclareSymbolFont{mathx}{U}{mathx}{m}{n}
\DeclareMathDelimiter{\vvvert} {0}{matha}{"7E}{mathx}{"17}
\DeclarePairedDelimiterX{\normiii}[1]{\vvvert}{\vvvert} {\ifblank{#1}{\:\cdot\:}{#1}}

\DeclarePairedDelimiter\paren\lparen\rparen
\DeclarePairedDelimiter\rb\lparen\rparen

\DeclarePairedDelimiterX\norm[1]\lVert\rVert{\ifblank{#1}{\placeholder}{#1}}
\DeclarePairedDelimiter\abs\lvert\rvert

\DeclarePairedDelimiterX\ip[2]{\langle}{\rangle}{#1 , #2}
\DeclarePairedDelimiterX\dual[2]{\langle}{\rangle}{#1 , #2}

\newcommand\qandq{\quad\text{and}\quad}

\DeclarePairedDelimiterX\braket[2]\langle\rangle{ #1 \delimsize\vert #2}
\DeclarePairedDelimiterX\ketbra[2]\vert\vert{ #1 \delimsize\rangle\delimsize\langle #2 }

\providecommand\given{}  
\newcommand{\SetSymbol}[1][]{ \nonscript\ #1\vert \allowbreak \nonscript\ \mathopen{} }
\DeclarePairedDelimiterX{\set}[1]\{\}{ \renewcommand\given{\SetSymbol[\delimsize]} #1 }

\def\SU{\operatorname{SU}}
\def\su{\mathfrak{su}}
\def\SO{\operatorname{SO}}
\def\so{\mathfrak{so}}

\newcommand\supind[1]{^{(#1)}}

\definecolor{darkblue}{rgb}{0.1, 0.1, 0.6}

\definecolor{0}{RGB}{102,102,102}
\definecolor{1}{RGB}{85,104,184}
\definecolor{2}{RGB}{218,34,34}
\definecolor{3}{RGB}{119,183,125}
\definecolor{4}{RGB}{231,140,53}
\definecolor{5}{RGB}{155,98,167}
\definecolor{6}{RGB}{78,121,197}
\definecolor{7}{RGB}{223,72,40}
\definecolor{8}{RGB}{140,188,104}
\definecolor{9}{RGB}{228,156,57}
\definecolor{10}{RGB}{140,78,153}
\definecolor{11}{RGB}{84,158,179}
\definecolor{12}{RGB}{82,26,19}
\definecolor{13}{RGB}{209,181,65}
\definecolor{14}{RGB}{228,99,45}
\definecolor{15}{RGB}{209,193,225}
\definecolor{16}{RGB}{89,165,169}
\definecolor{17}{RGB}{184,34,30}
\definecolor{18}{RGB}{221,170,60}
\definecolor{19}{RGB}{195,168,209}
\definecolor{20}{RGB}{166,190,84}
\definecolor{21}{RGB}{78,150,188}
\definecolor{22}{RGB}{230,121,50}
\definecolor{23}{RGB}{111,76,155}
\definecolor{24}{RGB}{149,33,27}
\definecolor{25}{RGB}{190,188,72}
\definecolor{26}{RGB}{181,143,194}
\definecolor{27}{RGB}{114,30,23}
\definecolor{28}{RGB}{167,120,180}
\definecolor{29}{RGB}{77,138,198}
\definecolor{30}{RGB}{105,177,144}
\definecolor{31}{RGB}{221,216,239}
\definecolor{32}{RGB}{96,89,169}
\definecolor{33}{RGB}{96,171,158}
\definecolor{34}{RGB}{232,236,251}

\newcommand{\ox}{\otimes}

\newcommand{\bra}[1]{\langle #1|}
\newcommand{\ket}[1]{|#1\rangle}

\newcommand{\emp}[2][b]{\if b#1\textbf{#2}\else\textit{#2}\fi}

\renewcommand{\qquad}{\hspace{60pt}}

\newcommand{\sref}[2]{\hyperref[#2]{#1 \ref*{#2}}}

\newcommand{\B}{\mathcal{B}}

\newcommand{\CC}{\mathbb{C}}

\renewcommand{\d}{\ \textrm{d}}

\renewcommand{\H}{\mathcal{H}}

\newcommand{\M}{\mathcal{M}}

\newcommand{\N}{N}

\renewcommand{\P}{\mathcal{P}}

\newcommand{\U}{\mathcal{U}}

\newcommand{\UUU}{\mathscr{U}}

\newcommand{\V}{\mathcal{V}}

\newcommand{\VVV}{\mathscr{V}}

\DeclareSymbolFont{usualmathcal}{OMS}{cmsy}{m}{n}
\DeclareSymbolFontAlphabet{\mathcal}{usualmathcal}

\begin{document}

\begin{center}{\Large \textbf{
Error bounds for Lie Group representations in quantum mechanics\\
}}\end{center}

\begin{center}
Lauritz van Luijk\textsuperscript{1$\star$},
Niklas Galke\textsuperscript{2},
Alexander Hahn\textsuperscript{3} and
Daniel Burgarth\textsuperscript{3}
\end{center}

\begin{center}
{\bf 1} Institut für Theoretische Physik, Leibniz Universität Hannover, Appelstraße 2, 30167 Hannover, Germany
\\
{\bf 2} F\'{\i}sica Te\`{o}rica: Informaci\'{o} i Fen\`{o}mens Qu\`{a}ntics, Departament de F\'{\i}sica, Universitat Aut\`{o}noma de Barcelona, 08193 Bellaterra, Spain
\\
{\bf 3} Center for Engineered Quantum Systems, Macquarie University, 2109 NSW, Australia
\\

${}^\star$ {\small \sf lauritz.vanluijk@itp.uni-hannover.de}
\end{center}

\begin{center}
\today
\end{center}

\section*{Abstract}
{\bf
We provide state-dependent error bounds for strongly continuous unitary representations of connected Lie groups.
That is, we bound the difference of two unitaries applied to a state in terms of the energy with respect to a reference Hamiltonian associated with the representation and a left-invariant metric distance on the group.
Our method works for any connected Lie group, and the metric is independent of the chosen representation.
The approach also applies to projective representations and allows us to provide bounds on the energy-constrained diamond norm distance of any suitably continuous channel representation of the group.
}

\vspace{10pt}
\noindent\rule{\textwidth}{1pt}
\tableofcontents\thispagestyle{fancy}
\noindent\rule{\textwidth}{1pt}
\vspace{10pt}
\markboth{}{}

\section{Introduction}

By providing a framework to describe continuous symmetries in physics, Lie groups are assigned an important role in theoretical and mathematical physics.
These symmetries are implemented on a physical system in terms of a representation of the group.
The representation theory of Lie groups plays a major role in many areas of physics:
For instance, in classical mechanics, it appears in the study of orbital angular momentum \cite{Thompson1994} and of the symplectic structure of the phase space \cite{Kim1991}.
In quantum theory, prominent examples are quantum control theory \cite{DAlessandro2007} or quantum field theory \cite{Kerf1997}.
Finally, in special and general relativity, it is essential to study the representation theory of the Lorentz group and the Poincar\'e group as the symmetry groups of Minkowski spacetime \cite{Carroll2019}.

A symmetry of a quantum mechanical system is an invertible transformation of the state space that doesn't lead to observable differences in the statistics.
By Wigner's theorem \cite{Wigner1931}, such symmetries are implemented by unitary (or anti-unitary) operators. 
Due to this, continuous quantum mechanical symmetries are described by \emph{projective unitary representations} of connected Lie groups. 
These are unitary operators $U_g$, whose multiplication recovers the group operation up to a phase. 
If the phase is trivial, one simply speaks of a \emph{unitary representation} or of a \emph{proper} unitary representation to emphasize.
In practice, \emph{projective} unitary representations often occur if one exponentiates a representation of a Lie algebra: 
A Lie algebra representation might not exponentiate to the group of interest, but rather a covering group \cite{Hall2015}. 
Take for example the spin-$\frac12$ representation of $\mathfrak{so}(3)$.
Conceptually, it often makes more sense to view such a representation as a projective representation, i.e.\ a ``representation  up to phase'', of the group at hand, instead of viewing it as a (proper) unitary representation of its cover.
In any case, the issue of additional phases only arises on the Hilbert space level. 
For the induced unitary quantum channels $\U_g=U_g(\placeholder)U_g^*$, the phases from $U_g$ and $U_g^*$ cancel each other ($U_g^*$ denoting the adjoint of $U_g$).

In practice, (projective) unitary representations of Lie groups might act on high-dimensional or even infinite-dimensional Hilbert spaces. This can be unfavorable, in particular when performing numerical simulations.
Therefore, a common practice is to instead study the (matrix) Lie group itself.
For example, an alternative for the infinite-dimensional time-evolution under a quadratic bosonic Hamiltonian (metaplectic group) can be achieved by exponentiating a corresponding finite-dimensional matrix, see \cref{sec:metaplectic}. 
However, a priori, the distance between two unitaries cannot directly be inferred through this procedure. 
Yet, this quantity is of particular interest with applications including quantum speed limits \cite{Deffner2017,Becker2021}, quantification of Trotter errors \cite{Childs2021}, and quantum channel capacities \cite{Winter2017}. 
Hence, the objective is to relate the distance between two unitary transformations to a suitable notion of distance on the underlying group.
In this paper, we do this by providing explicit error bounds on the distance between two unitaries $U_g$ and $U_h$ in terms of a metric $d(g,h)$ on the Lie group $G$.

In the generic case of a unitary representation acting on a potentially infinite-dimen\-sional Hilbert space $\H$, the operator norm distance between two unitaries might attain its maximum of $2$, even if the unitaries are -- intuitively speaking -- ``close'' to each other, i.e.\ the representation is not uniformly continuous. 
See \cite[Sec.~3]{Ichinose2004} for an example. For this reason, it is necessary to quantify the distance in a ``weaker'' way.
This can be achieved by state-dependent error bounds.
These are continuity bounds on the representation w.r.t.\ the strong operator topology, which is a weaker topology than the uniform (or operator norm) topology.
In the picture of quantum channels, these strong error bounds translate to bounds on the \emph{energy-constrained diamond norm} \cite{shirokov}, where only input states up to a certain energy $E$ with respect to a suitable reference Hamiltonian are considered.
It will turn out that a natural and useful reference Hamiltonian is the \emph{Nelson Laplacian} $\Nelson$.
This is a well-studied operator which plays a central role in infinite-dimensional representations of Lie algebras \cite{nelson1959analytic}. 
Using the state-dependent error measures, we find bounds for the unitary representations $U_g$ and $U_h$
\begin{equation*}
        \norm{(U_g-U_h)\psi } \leq \sqrt{\ip\psi{\Nelson\psi}} \, d(g,h)\quad \text{for all $\psi\in\H$ and $g,h\in G$}
   \end{equation*}
with metric $d(g,h)$ defined in \cref{eq:metric_def} below and
\begin{equation*}
        \norm{\U_g(\rho) - \U_h(\rho)}_1 \le 2 \sqrt{\tr[\rho\Nelson]}\,d(g,h)
\end{equation*}
for the unitary channels $\U_g$ and $\U_h$.
The result in the channel case can be stated in terms of bounds on the \emph{energy-constrained diamond norm distance}:
$$\norm{\U_g - \U_h}^{\Nelson, E}_{\diamond}\le 2 \sqrt{E}\,d(g,h).$$ Our method is general, so that these bounds hold \emph{for all} (connected) Lie groups. Therefore, the bounds can be used directly for any computation involving unitary or unitary channel representations of Lie groups. Furthermore, we extend our result to projective unitary representations at the end of \cref{sec:projective}.
To emphasize the difference, we will sometimes use the term \emph{proper} unitary representations if a unitary representation is explicitly \emph{not} a projective one.
For definitions of the different types of representations, see \cref{sec:main}.

The remaining paper is structured as follows. \cref{sec:preliminaries} collects some preliminaries and useful results. In particular, \cref{sec:Nelson_metric} introduces a metric with which we endow the Lie group. It is followed by a discussion about $\Ad$-invariant inner products in \cref{sec:riem}. The actual strong error bounds are presented in \cref{sec:main}. Here, \cref{sec:unitary_reps} deals with proper unitary representations. These bounds are extended in \cref{sec:projective} to representations by unitary channels. However, we are also able to give an error bound for the projective unitary representation implementing the channels if the group elements are sufficiently close to each other. The abstract theory from \cref{sec:preliminaries} and \ref{sec:main} is complemented by a case study of six physically relevant examples in \cref{sec:examples}. 
Specifically, we consider the special unitary group, whose representations describe quantum spin systems in \cref{sec:SU(2)}, the special orthogonal group which plays a role for free fermion models in \cref{sec:free_fermion}, displacement operators in \cref{sec:displacement}, the metaplectic group with applications in quantum optics in \cref{sec:metaplectic}, the group $\mathrm{SU}(1,1)$ for interferometry in \cref{sec:su11} and the Lorentz group with its application to spinless particles in relativistic quantum mechanics in \cref{sec:Lorentz}. 
The proofs of our results are presented in \cref{sec:proofs}. Finally, we conclude in \cref{sec:conclusion}.
In the appendix, we explicitly present the calculation of the Nelson Laplacian for the metaplectic representation (\cref{appendix:Nelson_symplectic}) and the fermionic linear optics (FLO) representation (\cref{appendix:Nelson_SO}).

\section{Preliminaries}\label{sec:preliminaries}

In this section, we define some basic notions that we will use throughout the paper. In particular, we introduce a left-invariant metric on the Lie group and discuss some of its properties. Furthermore, we discuss the special case when the Lie algebra has an $\Ad$-invariant inner product.

Throughout the paper, the complex conjugate of a number $z\in\CC$ will be denoted $\bar z$. 
Furthermore, the hermitian adjoint of a complex matrix $A$ and the adjoint operator of an operator on an abstract Hilbert space will be denoted $A^*$.

\subsection{The metric}\label{sec:Nelson_metric}

Let $G$ be a connected Lie group and let $\lie$ be its Lie algebra. 
We will always assume $\lie$ to be equipped with an inner product $\ip\placeholder\placeholder_\lie$, i.e., $\ip\placeholder\placeholder_\lie:\lie\times\lie\to\RR$ is a positive definitive bilinear form.
While the specific choice of the inner product does not matter for the abstract theory, it is important for the interpretation of the bounds and can even be used to account for cases where some of the generators are harder to implement than others.
This makes the freedom of choice of the metric a feature of our results (also see Lemma~\ref{thm:lemma_nelson}\ref{it:equivalent_NL}).
In cases of matrix groups $G\subset \mathrm{Gl}(n,\RR)$ (or $G\subset\mathrm{Gl}(n,\CC)$) we will typically use the inner product $(X,Y)\mapsto \tr X^\top Y$ (respectively, $(X,Y)\mapsto \tr X^* Y$).
Notice that the Cartan-Killing form could, in principle, also be used if it is positive definite (or negative definite, depending on the convention).
However, this requires compactness and a semisimple Lie algebra, which is not always the case.

\begin{defin}[Left-invariant metric]
	We define a left-invariant metric $d:G\times G\to\RR_+$ as follows,
\begin{equation}\label{eq:metric_def}
    d(g,h) = \inf \set[\bigg]{\,\sum_{j=1}^n\, \norm{Y_j}_\lie \given n\in\NN,\ Y_1,\dots,Y_n\in\lie : g = h\,\rme^{Y_1} \cdots \rme^{Y_n}  }.
\end{equation}
\end{defin}

Left-invariance of the metric $d$ means that 
\begin{equation}\label{eq:left_inv}
    d(gh_1,gh_2) = d(h_1,h_2),\quad g,h_1,h_2\in G.
\end{equation}
Since any element of a connected Lie group is a finite product of exponentials \cite[Thm.~IV.1]{Chevalley1946}, the metric $d(g,h)$ is always finite. In order to demonstrate that our definition of $d(g,h)$ is indeed reasonable, we prove the following lemma. All proofs of the subsequent results are presented in \cref{sec:proofs}.

\begin{lem}\label{thm:lemma_metric}
    $d$ is a left-invariant metric on $G$ and the induced metric topology agrees with the topology of $G$.
\end{lem}

Notice that \cref{thm:lemma_metric} holds for compact \emph{and} non-compact Lie groups, and we will see examples for both cases in \cref{sec:examples}.
For a discussion on the topology of group manifolds, see, for instance,~\cite[Ch.~2.2,~11]{Gilmore2008}\@.
If $g$ and $h$ are close enough for $g^{-1}h$ to have a logarithm in $\lie$, the metric is always bounded by
\begin{equation}\label{eq:metric_to_norm}
    d(g,h) \leq \norm{\log(g^{-1}h)}_\lie.
\end{equation}
To be precise, the estimate $d(g,h)\le \norm X_\lie$ holds for all $X\in\lie$ such that $e^X=g^{-1}h$.
If the inner product $\ip\placeholder\placeholder_\lie$ is $\Ad$-invariant, then we even have equality in \eqref{eq:metric_to_norm}, and the metric $d$ is equal to the Riemannian distance with respect to the corresponding bi-invariant Riemannian geometry.
Here, $\Ad$-invariance means that $\ip{\Ad_g X}{\Ad_g Y}{}{}_\lie=\ip{X}{Y}{}{}_\lie$ for all $g\in G$.
Notice that if $G$ is a compact Lie group, then its Lie algebra $\lie$ always admits an $\Ad$-invariant inner product as a real vector space~\cite[Prop.~4.24]{Knapp1996}\@.

\subsection{Lie algebras with $\Ad$-invariant inner product}\label{sec:riem}

An inner product on the Lie algebra induces a left-invariant Riemannian geometry on the Lie group. If the inner product is $\Ad$-invariant (see below), then this geometry is also right-invariant, and its geodesics are exponentials.

Using the differential of the left translation $\ell_g : G\to G$, $h\mapsto gh$, the inner product $\ip\placeholder\placeholder_\lie$ on the Lie algebra $\lie$ can be extended to a left-invariant Riemannian metric on $G$. 
For this, note that the Lie algebra $\lie$ is isomorphic to the tangent space $T_1G$ at the neutral element $1\in G$.
We can now use that the differential $\diff \ell_g |_1:T_1 G\to T_g G$ is an isomorphism between the tangent space at the neutral element and the tangent space at $g\in G$. 
We abuse notation by writing $gX$ for $\diff\ell_g(X)$. $Xg$ is defined analogously and we have $\Ad_g(X) = gXg^{-1}$. 
The Riemannian metric at $g\in G$ is then simply defined by
\begin{equation}\label{eq:}
    \ip XY_g = \ip{g^{-1}X}{g^{-1}Y}_\lie,\quad \text{$X,Y\in T_gG$}.
\end{equation}
We assume throughout this section that $\ip\placeholder\placeholder_\lie$ is $\Ad$-invariant, i.e.\ $\ip{gXg^{-1}}{gYg^{-1}}_\lie = \\\ip XY_\lie$ for all $g\in G$, $X,Y\in\lie$.  
An $\Ad$-invariant inner-product exists always on compact as well as abelian Lie groups \cite{Milnor1976}.
The Riemannian geometry on $G$ induced by an $\Ad$-invariant inner-product on $\lie$ is bi-invariant. 
That is, the differentials of left and right translations are isometric.

Recall that the Riemannian length of a curve $\gamma:[0,T]\to G$ is defined as
\begin{equation}
    \Length{\gamma} 
    = \int_0^T \norm{\dot \gamma(t)}_{\gamma(t)} \,\diff t
\end{equation}
and the Riemannian distance of two elements $g,h\in G$ is
\begin{equation}
    d_{\textup{Riem}}(g,h) = \inf_\gamma \Length{\gamma},
\end{equation}
where the infimum is over all curves $\gamma$ connecting $g$ and $h$.

\begin{prop}\label{thm:riemannian_metric}
    Assume that $\ip\placeholder\placeholder_\lie$ is an $\Ad$-invariant inner product on $\lie$ and equip $G$ with the induced bi-invariant Riemannian geometry.
    Then the metric $d$, defined in \cref{eq:metric_def}, is equal to the Riemannian distance, i.e.\
    \begin{equation}\label{eq:riemannian_metric}
        d(g,h) = d_{\textup{Riem}}(g,h).
    \end{equation}
    In this case, let $V \subset G$ be a neighborhood of the identity, such that $\log = (\exp)^{-1} :V\to \lie$ is well-defined, then
    \begin{equation}\label{eq:metric_log}
        g^{-1}h \in V \implies d(g,h) = \norm{\log (g^{-1} h)}_\lie.
    \end{equation}
\end{prop}

\section{Main results}\label{sec:main}

After establishing the basic concepts, we are now ready to state the main results, i.e.\ we present a strong error bound on unitary representations of Lie groups. Afterwards, this bound is extended to unitary channels as well as projective unitary representations.

\subsection{Proper unitary representations}\label{sec:unitary_reps}

Here, we consider the case of a proper unitary representation.
As in \cref{sec:preliminaries}, consider a Lie group $G$ and an inner product $\ip\placeholder\placeholder_\lie$ on its Lie algebra $\lie$.
Let $(U,\H)$ be a strongly continuous unitary representation of $G$. That is, $\H$ is a separable Hilbert space and
\begin{equation}
    U : G \to \mathrm{U}(\H)\label{eq:unitary_rep}
\end{equation}
is a continuous group homomorphism, where $\mathrm U(\H)$ is the group of unitary operators on $\H$ endowed with the strong operator topology.
In particular, $U_gU_h = U_{gh}$ for any $g,h\in G$.
In fact, this is what we mean by a \emph{proper} unitary representation if we want to emphasize that the group multiplication is respected exactly by the unitary representation without an additional phase factor.
We denote by $\dU$ the induced representation of the Lie algebra in terms of self-adjoint operators, i.e.\
\begin{equation}
    U(\rme^{X}) = \rme^{-\rmi\dU(X)} ,\quad \text{for all } X\in\lie.
\end{equation}
There is a common invariant domain $\mathcal D \subset \H$ for all $A(X)$, on which they satisfy  \cite{nelson1959analytic}
\begin{equation}\label{eq:LieAlgRep}
    [A(X),A(Y)] = \rmi A([X,Y])\quad\text{for all $X,Y\in\lie$}.
\end{equation}
Notice that $A(X)$ is self-adjoint while the infinitesimal generator $X$ typically is a (real) skew-symmetric matrix.
This is also the reason for the appearance of the imaginary unit on the right-hand side of \cref{eq:LieAlgRep}.
The \emph{Nelson Laplacian} of the representation $(U,\H)$ is the self-adjoint unbounded operator
\begin{equation}\label{eq:nelson_lap}
    \Nelson = \sum_j [\dU(X_j)]^2,
\end{equation}
where $\set{X_j}$ is any orthonormal basis with respect to the inner product $\ip\placeholder\placeholder_\lie$ \cite{nelson1959analytic}.
All generators $A(X)$ are relatively bounded with respect to the Nelson Laplacian and have relative bound $0$. In particular, $\dom \Nelson \subset \dom A(X)$.
The Nelson Laplacian only depends on the inner product but not on the choice of orthonormal basis, see \cref{thm:lemma_nelson}.
By construction, the Nelson Laplacian is always positive definite, $\Nelson\geq0$.

\begin{lem}\label{thm:lemma_nelson}
	\begin{enumerate}[(1)]
        \item\label{it:nelson_basis_dep}
            The Nelson Laplacian $\Nelson$ is independent of the chosen orthonormal basis.
        \item\label{it:beta_ineq}
            For all $X\in\lie$ and $\psi\in\dom{\sqrt\Nelson}$, one has $\psi\in\dom A(X)$ and
            \begin{equation}\label{eq:beta_ineq}
                \norm{\dU(X)\psi} \leq \norm{X}_\lie \sqrt{\ip\psi{\Nelson\psi}}.
            \end{equation}
        \item\label{it:equivalent_NL}
            If $\ip\placeholder\placeholder_{\lie}'$ is another inner product on $\lie$ and if $\Nelson'$ is the associated Nelson Laplacian then
            \begin{equation}\label{eq:equivalent_NL}
                c \Nelson' \le \Nelson \le C\Nelson',
            \end{equation}
            or, equivalently, $C^{-1}\Nelson\le \Nelson'\le c^{-1}\Nelson$,
            where $c,C>0$ are constants such that $c\ip XX_{\lie}\le\ip XX_{\lie}'\le C \ip XX_{\lie}$ for all $X\in \lie$.
    \end{enumerate}
\end{lem}

For projective unitary representations, the Nelson Laplacian can be defined in a similar way as in \cref{eq:nelson_lap}. In this case, all its relevant properties from \cref{thm:lemma_nelson} still hold true, see \cref{sec:projective}.
As a simple example for \cref{eq:beta_ineq}, consider the case $G= \RR$ with the representation $U(t) = \rme^{-\rmi tH}$ for some Hamiltonian $H$.
For the inner product on the Lie algebra $\lie =\RR$ we choose the ordinary multiplication of real numbers, so $\norm\placeholder_\lie$ is just the absolute value.
Then the bound in \cref{eq:beta_ineq} trivially is an equality $\norm{tH\psi} = \abs t\sqrt{\ip\psi{H^2\psi}}$. Note that $\Nelson = H^2$ in this case.

In concrete examples, the inner product can be adjusted so that either one obtains a specific Nelson Laplacian (like the harmonic oscillator) or to account for situations where different directions in $\lie$ are ``more expensive'' than others.
We use the Nelson Laplacian in order to find a bound for the distance between two elements from a proper unitary representation.

\begin{thm}\label{thm:main}
    Let $G$ be a connected Lie group and let $\ip\placeholder\placeholder_\lie$ be an inner product on its Lie algebra $\lie$. 
    Then for any strongly continuous unitary representation $(U,\H)$ with associated Nelson Laplacian $\Nelson$ one has
    \begin{equation}\label{eq:main_bound}
        \norm{(U_g-U_h)\psi } \leq \sqrt{\ip\psi{\Nelson\psi}} \, d(g,h)\quad \text{for all $\psi\in\H$ and $g,h\in G$},
    \end{equation}
    where $d$ is the metric defined by \cref{eq:metric_def}.
\end{thm}

In \cref{eq:main_bound}, we use the convention that $\ip\psi{\Nelson\psi} \coloneqq\norm{\sqrt\Nelson \psi}^2$ if $\psi\in\dom{\sqrt\Nelson}$ and $\ip\psi{\Delta\psi}=\infty$ else.
Note that this bound is universal in the sense that the metric is independent of the chosen representation $U$ and that the metric itself only depends on the inner product on $\lie$ and, of course, the group $G$.

\begin{prop}\label{thm:corollary}
    Under the assumptions of \cref{thm:main},
	let $K$ be a positive operator such that for all $X\in\lie$ $\dom{\sqrt{K}}\subset \dom{\dU(X)}$ and
	\begin{equation}\label{eq:bertaN}
	    \Vert \dU(X)\psi\Vert\leq \Vert X\Vert_\lie \sqrt{\ip\psi{K\psi}}\quad \text{ for all $\psi\in\dom{\sqrt{K}}$}.
	\end{equation}
	Then the bound \cref{eq:main_bound} also holds for $K$, i.e.\
	\begin{equation}
		\norm{(U_g-U_h)\psi } \leq \sqrt{\ip\psi{K\psi}} \, d(g,h)\quad \text{for all $\psi\in\H$ and $g,h\in G$}.
	\end{equation}
\end{prop}

This is, of course, only advantageous if the operator $K$ is smaller than the Nelson Laplacian.
Note that condition \eqref{eq:bertaN} can be written as the operator inequality $[A(X)]^2 \le \norm{X}_\lie^2\,K$ for all $X\in\lie$.

\subsection{Projective representations and estimates on unitary channels}\label{sec:projective}

In many cases, describing symmetries of quantum systems by proper unitary representations is too restrictive.
Since the measurement statistics are insensitive to global phases, two unitaries, which differ only by a phase, cannot be distinguished by measuring the isolated system.
More concretely, what really matters is the implemented unitary channel $\U:\rho\mapsto U \rho U^*$ with $\rho \in \states(\H)$, where $\states(\H)$ denotes the set of density operators on $\H$.
Recall that a density operator is a positive operator with unit trace.
$\U$ is an example of a \emph{quantum channel} \cite{Wolf2012}. Quantum channels are completely positive trace preserving (CPTP) maps on the Banach space $\traceclass$ of trace class operators on $\H$.
It is noteworthy that every quantum channel, which has an inverse channel, is already a unitary channel. This follows from Wigner's Theorem \cite{Wigner1931} and the fact that conjugation by anti-unitary operators is not completely positive.
\begin{defin}
    A \emph{strongly continuous unitary channel representation} of a connected Lie group $G$ is a tuple $(\U,\H)$ where $\H$ is a Hilbert space and $g\mapsto \U_g$ assigns a unitary channel to each group element such that
    \begin{enumerate}[label=\arabic*)]
    \item
        $\U_1 = \id$ where $1\in G$ is the neutral element of $G$ and $\id$ is the identity channel, i.e.\ $\id(\rho)=\rho$ for all $\rho\in\traceclass$.
    \item\label{item:rep}
        For all $g,h\in G$ we have that $\U_g\U_h = \U_{gh}$.
    \item
        For every $\rho\in\traceclass$ the map $g\mapsto\U_g(\rho)$ is continuous with respect to the trace norm $\norm{\rho}_1=\tr\sqrt{\rho^*\rho}$.
\end{enumerate}
\end{defin}

Henceforth, $(\U,\H)$ denotes a strongly continuous unitary channel representation.
We will now investigate the relationship between strongly continuous unitary channel representations and projective unitary representations. This allows us to define the Nelson Laplacian for a general strongly continuous unitary channel representation.

It is clear that every strongly continuous unitary representation $(U,\H)$ defines a strongly continuous unitary channel representation $(\U,\H)$, but the converse is not true.
In fact, a strongly continuous unitary channel representation $(\U,\H)$ only gives rise to a projective unitary representation in general.

This projective representation is obtained from $(\U,\H)$ by any choice of implementing unitaries $U_g$, i.e.\ unitaries such that $\U_g(\rho) = U_g\rho U_g^*$ for all $g\in G$.
These unitaries automatically satisfy
\begin{equation}
    U_g U_h = \zeta(g,h) U_{gh}\quad \text{for all $g,h\in G$}
\end{equation}
for some phase factors $\zeta(g,h)\in \mathrm{U}(1)$.
These phase factors distinguish a projective from a proper unitary representation.
According to \cite[Thm.~1.1]{Bargmann1954}, it is always possible to find implementing unitaries such that there is a neighborhood $\UUU$ of $1\in G$, within which the following equivalent conditions hold
\begin{enumerate}[label=\alph*)]
    \item
         $g\mapsto U_g\psi$ is continuous on $\UUU$ for each $\psi\in\H$,
     \item
         $\zeta$ is continuous on $\UUU\times\UUU$.
\end{enumerate}
A detailed treatment of projective unitary representations of topological groups is given in Ref.~\cite{Bargmann1954}.
For our purposes, we can summarise as follows: For any strongly continuous unitary channel representation $(\U,\H)$ of $G$, there is a projective unitary representation $(U,\H)$, which implements $(\U,\H)$ and is strongly continuous in a neighborhood of the neutral element $1\in G$. 
Conversely, any projective unitary representation $(U,\H)$, which is strongly continuous near the neutral element $1\in G$, determines a strongly continuous unitary channel representation $(\U,\H)$ by $\U_g(\rho) = U_g\rho U_g^*$.
We will use such a projective representation to define the Nelson Laplacian of $(\U,\H)$.

We start with one-parameter groups:
Let $\tau > 0$ and let $U: (-\tau,\tau)\to\B(\H)$ be a map with the following properties
\begin{enumerate}[label=\roman*)]
    \item
    $U(t)$ is a unitary for each $-\tau < t < \tau$,
    \item 
    $U(0) = \1$ and $U$ is strongly continuous in $0$, and
    \item
    for all $-\tau < t,s < \tau$ such that $-\tau <t+s<\tau$ it holds that $U(t+s) = U(t)U(s)$.
\end{enumerate}
Then we can extend $U$ to be a strongly continuous one-parameter unitary group by setting $U(s) = U(t_1)\cdots U(t_N)$ for $\RR\ni s = t_1 + \cdots + t_N$, with $t_k\in (-\tau,\tau)$. It is straightforward to check that this does not depend on the chosen decomposition.
By the Stone-von Neumann theorem \cite[Thm.~VIII.12]{reed2012methods}, there exists a generator $A$ such that $U(t) = \rme^{-\rmi t A}$ for $t \in (-\tau,\tau)$.

Now, let $U$ be a projective unitary representation of $G$, which is strongly continuous in a neighborhood of the neutral element $1\in G$.
By \cite[Lem.~4.4]{Bargmann1954}, $t\mapsto U_{\rme^{tX}}$ may be restricted to a suitable interval, such that it fulfills the above properties.
Hence, there exists a generator $\dU(X)$ such that $U_{\rme^{tX}} = \rme^{-\rmi t\dU(X)}$ for $t$ small enough.
\begin{rem}\label{rem:projgen}
    Consider a strongly continuous unitary channel representation.
    The results \cite[Thm.~3.3, Thm.~5.1]{Bargmann1954} tell us that there exists a strongly continuous projective unitary representation $(V,\H)$ of the universal covering group $\covering$ (see below or \cite[Sec.~3e]{Bargmann1954}) of $G$ such that $V_{\covering[\rme]^{tX}} = U_{\rme^{tX}}$ for all $X\in\lie$ and $t$ small enough.
    Here, $\covering[\rme]^{tX} = \exp_{\covering}(tX)$ denotes the exponential map of $\covering$.
    The corresponding central extension $\centext$ (see \cite[Sec.~5]{Bargmann1954}) of $\covering $, whose Lie algebra is $\RR\oplus\lie$ as a vector space, then admits a strongly continuous unitary representation $W$ with $W_{\rme^{t(0\oplus X)}} = V_{\covering[\rme]^{tX}} = U_{\rme^{tX}}$, again for $t$ small enough.
    If $\H$ is finite-dimensional then $V$ can even be chosen to be a proper unitary representation of $\covering$ \cite[Sect.~3b]{Bargmann1954} (and there is no need for a central extension).
\end{rem}

Combining the above one gets the following characterization of the Lie algebra ``representation'' induced by unitary channel representation:

\begin{lem}\label{lem:projgen}
    If $(U,\H)$ is a projective unitary representation, which is continuous near $1\in G$, then there exists a corresponding representation $\dU$ of $\lie$ by self-adjoint operators such that for $t$ small enough
    \begin{align}
        \rme^{-\rmi t\dU(X)} = U_{\rme^{tX}},\quad \text{if }\, t\approx0.
    \end{align}
    The generators $A(X)$ have a common invariant domain on which they satisfy
    \begin{equation}\label{eq:LieAlgRepProjective}
        [A(X),A(Y)] = \rmi A([X,Y]) + \rmi\lambda(X,Y)\1,
    \end{equation}
    where $\lambda :\lie\times\lie\to \RR$ is an infinitesimal version of the map $\zeta$, see \cite[Sec.~4f.]{Bargmann1954}.
\end{lem}

Note that \cref{eq:LieAlgRepProjective} is the analog of \cref{eq:LieAlgRep} for projective representations.
We can now define the Nelson Laplacian $\Nelson$ for any strongly continuous unitary channel representation $(\U,\H)$ of a Lie group $G$ in terms of an implementing projective representation, which is strongly continuous near $1\in G$ as before:
We pick an ONB $\set{X_j}$ of $\lie$ and set $\Nelson = \sum_j A(X_j)^2$.
\cref{rem:projgen} makes sure that the functional analytical properties of $\Nelson$ and its relation to the generators $\dU(X)$ are as before.
The formula $\norm{\dU(X)\psi}\le \norm{X}_\lie \sqrt{\ip\psi {\Nelson\psi}}$ also generalises for all $\psi\in\dom\sqrt\Nelson$ (see proof of Theorem~\hyperlink{prf:thm_channel_bound}{\ref{thm:channel_bound}}).

\begin{rem}
In fact, we can choose the inner product on $\RR\oplus\lie$ such that any orthonormal basis of $\lie$ can be completed to one of the extended Lie algebra.
The remaining basis vector will be represented by a multiple of the identity operator.
Hence, the complete Nelson Laplacian of $\centext$ differs from $\Nelson$ only by a multiple of the identity.
Lastly, $\Nelson$ is again independent of the choice of orthonormal basis.
This is because orthogonal transformations of $(\lie, \ip{\placeholder}{\placeholder}_\lie)$ can be extended to orthogonal transformations of $\RR\oplus\lie$.
\end{rem}

This can be used to bound the distance of the unitary channels $\U_g$ in the \emph{energy-constrained diamond norm (ECD norm)} \cite{shirokov, Winter2017}.

\begin{defin}
    The ECD norm (with respect to $\Nelson$) of a linear map $\Phi$ on $\traceclass$ is defined for any $E > \inf \mathrm{spec}(\Nelson)$ as
    \begin{equation}
        \norm{\Phi}^{\Nelson, E}_\diamond = \sup_{\substack{\rho\in\states(\H\otimes\H')\\ \tr{\rho\Nelson\otimes \1}\le E}} \norm{\Phi\otimes\id_{\mathfrak{T}(\H')} (\rho)}_1,
    \end{equation}
    where the supremum is over all auxiliary systems $\H'$.
\end{defin}
We are now ready to state an analogous result as the bound in \cref{thm:main} for strongly continuous unitary channel representations.

\begin{thm}\label{thm:channel_bound}
    Let $(\U,\H)$ be a strongly continuous unitary channel representation of a connected Lie group $G$.
    Then it holds that
    \begin{equation}
        \norm{\U_g - \U_h}^{\Nelson, E}_{\diamond} \le 2 \sqrt{E}\,d(g,h),
    \end{equation}
    where $\Nelson$ is the Nelson Laplacian of $(\U,\H)$ and $E > \inf \Spec(\Nelson)$.
\end{thm}
In particular, this result implies
\begin{equation}
    \norm{\U_g(\rho) - \U_h(\rho) }_1 \leq 2 \sqrt{\tr\rho \Nelson} \, d(g,h),
\end{equation}
for all density operators $\rho \in\states(\H)$.
Analogously to \cref{thm:corollary}, it is possible to obtain better bounds if one finds a positive operator $K\leq\Nelson$ such that \cref{eq:bertaN} holds.

The proof of \cref{{thm:channel_bound}} relies on \cite[Thm.~1]{Becker2021} stating that the ECD norm distance of two unitary channels can be computed by only considering pure states on $\H$.
However, the mentioned result restricts the dimension of $\H$ to be strictly greater than $2$.
This is due to the geometry of the (pure) state space of a qubit, which prevents the proof of \cite[Thm.~1]{Becker2021} to work for $\dim(\H) = 2$.
Nevertheless, the following result shows that the assertion is still true in this case.
We thank A.\ Winter for suggesting the main idea of the proof.

\begin{lem}\label{lem:dim2}
    Let $\U,\V$ be unitary channels implemented by unitaries $U,V$, respectively.
    Let $H$ be a Hamiltonian and $E > \inf \Spec(H)$.
    Then
    $$\norm{\U -\V}^{H, E}_\diamond = \sup_{\substack{\norm{\psi} = 1\\
    \ip\psi{H\psi}\le E}} \norm*{\rb*{\U - \V}(\ketbra\psi\psi)}_1.$$
    In particular, this holds for $\dim(\H) = 2$.
\end{lem}

A common situation in quantum mechanics is that a symmetry group $G$ is represented on a quantum system by a unitary channel (or projective unitary) representation, which comes from a proper unitary representation $(\tilde U,\H)$ of a covering group $\tilde G$.
Examples of this situation include the half-integer representations of $\mathrm{SU}(2)$, which is the double cover of $\mathrm{SO}(3)$, the metaplectic representation, and the half-integer representations of the Lorentz group.

We briefly recall the definition of covering groups for the reader's convenience:
A Lie group $\covering$ together with a smooth homomorphism $\pi:\covering\to G$ is a \emph{covering group} of $G$ if the differential $\diff \pi :\tilde\lie\to \lie$ is an isomorphism.
This is equivalent to the existence of a neighborhood $\VVV$ of $1\in G$ whose pre-image $\pi^{-1}(\VVV)$ is a disjoint union of open sets, each of which is diffeomorphic onto $\VVV$ via the restriction of $\pi$.
There always exists a unique (up to isomorphism) simply connected covering group $\covering$ \cite{Taylor1954}. This covering group is called the \emph{universal covering group} of $G$.

In the situation described above, where a projective representation comes from a proper representation of a covering group, one could apply the bounds on the unitary channels in \cref{thm:channel_bound}.
In particular, this is always the case if the Hilbert space $\H$ is finite-dimensional (see \cref{rem:projgen}).
However, sometimes it is actually useful to have strong error bounds on the unitaries themselves.
A simple workaround would be to apply \cref{thm:main} to the representation $(\tilde U,\H)$ of the covering group $\tilde G$.
Unfortunately, this approach is usually not applicable in practice because the covering group is often an abstract object and might not have a realization in terms of matrices. For instance, this is the case for the metaplectic group and the double cover of the Lorentz group.
Nevertheless, we can still obtain direct bounds for sufficiently close group elements in $\tilde G$ in terms of the metric on $G$. This is due to the fact that $\tilde G$ and $G$ are isomorphic locally.

\begin{prop}\label{thm:covering}
    Let $\tilde G$ be a covering group of $G$ and let $\ip\placeholder\placeholder_\lie$ be an inner-product on $\lie \equiv\tilde\lie$. 
    Consider the left-invariant metrics $d$ and  $\tilde d$ on $G$ and $\tilde G$, respectively, defined as in \cref{eq:metric_def}.
    Then the metrics $d$ and $\tilde d$ are equal locally, in the sense that $\tilde d(\tilde g,\tilde h) = d(g,h)$ whenever $\tilde g$ and $\tilde h$ as well as $g =\pi(\tilde g)$ and $h = \pi(\tilde h)$ are sufficiently close to each other. Here, $\pi:\tilde G\to G$ denotes the covering homomorphism.
\end{prop}

As an immediate application of this result, we get strong error bounds on the Hilbert space level for strongly continuous unitary representations of covering groups: 

Let $(\tilde U,\H)$ be an irreducible strongly continuous unitary representation of a covering group $\tilde G$ of $G$.
By choosing some arbitrary section $s:G\to \tilde G$, i.e. a map with $\pi \circ s =\id_G$, which is continuous near the neutral element, one can view $(\tilde U,\H)$ as a strongly continuous projective unitary representation of $G$.
We now get the following bounds:
\begin{equation}
    \left\Vert U_{g}\psi - U_{h} \psi \right\Vert \leq \sqrt{\ip\psi{\Nelson\psi}}\, d(g,h)
    \quad \text{if $g\approx h$}.\label{eq:projective_bound}
\end{equation}
Note that by our assumptions, the Nelson Laplacian does not depend on whether we view $(\tilde U,\H)$ as a proper representation of $\tilde G$ or as a projective representation of $G$.
The irreducibility is not essential, we only need to guarantee that the representation of $\tilde G$ maps $\pi^{-1}(1)$ to multiples of the identity $\1$.

In summary, \cref{eq:projective_bound} establishes a way to obtain error bounds even if we only have a projective unitary representation $(\tilde U,\H)$ of $G$ and not a proper unitary representation. 
The idea is to shift to the universal covering group $\tilde G$, on which $(\tilde U,\H)$ is a proper unitary representation. 
This works if the compared group elements $g$ and $h$ are sufficiently close to each other since $G$ and $\tilde G$ are locally isomorphic.
In this case, we can apply \cref{thm:main} to obtain error bounds in $\tilde G$. Reverting the local isomorphism allows us to shift these bounds back to $G$ and finally yields \cref{eq:projective_bound}.

\section{Examples and applications}\label{sec:examples}

In this section, we apply our bounds from \cref{sec:main} to some examples of physically important Lie groups. In particular, we study
\begin{enumerate}
	\item
		spin-$j$ systems: $\mathrm{SU}(2)$ in the spin representation (\cref{sec:SU(2)}),
	\item 
	    free fermion models: $\mathrm{SO}(2m)$ in the fermionic linear optics (FLO) representation (\cref{sec:free_fermion}),
	\item 
	    displacement operators: $\RR^{2m}$ in the Schr\"odinger representation (\cref{sec:displacement}),
	\item 
	    quasi-free bosonic transformations: $\mathrm{Sp}(2m,\RR)$ in the metaplectic representation (\cref{sec:metaplectic}),
	\item
	    $\mathrm{SU}(1,1)$ interferometry: $\mathrm{SU}(1,1)$ in the two-mode bosonic representation (\cref{sec:su11}) and
	\item
		 relativistic quantum mechanics: the scalar representation of the Lorentz group $\mathrm{SO}^+(1,3)$  (\cref{sec:Lorentz}).
\end{enumerate}
We compare our bounds to bounds in the literature, if there exist any. Apart from providing a framework that is valid for all connected Lie groups, we also find that our result yields tighter bounds in some cases.
Thus, we will use the convention that generators on the level of the Lie algebra are skew-symmetric, whereas generators on the Hilbert space level are self-adjoint.

\subsection{Spin representations of $\mathrm{SU}(2)$}\label{sec:SU(2)}

We consider the Lie group $G=\mathrm{SU}(2)$, i.e.\ the group of $2\times 2$ unitary matrices with determinant $1$. 
The Lie algebra $\mathfrak{su}(2)$ of trace-less skew-hermitian $2\times 2$ matrices will be equipped with the Frobenius inner product $\ip XY = \tr[X^*Y]$, which is $\Ad$-invariant.
Then, an orthonormal basis is given by $X_j = \frac \rmi2\sigma_j$, $j=1,2,3$, where $\sigma_j$ are the Pauli matrices.
The well-known commutation relations are $[X_i,X_j] = \varepsilon_{ijk} X_k$.
Topologically, $\mathrm{SU}(2)$ is the $3$-sphere $\mathbb{S}^3$. Since the inner product above is $\Ad$-invariant, we know that the metric on $\mathrm{SU}(2)$ is just the Riemannian distance on $\mathbb{S}^3$.
The Riemannian distance $d_{\mathrm{Riem}}(e^{-\rmi\alpha X},1)$ is $\norm{\log \rme^{-\rmi\alpha X}}_2 =\abs\alpha$ if $\alpha\in [-\pi,\pi]$ and $\norm{X}_2=1$.
More generally, we have $d_{\mathrm{Riem}}(g,1) = \norm{\log(g)}_2$.

We consider the spin-$j$ representation $U\up j : \mathrm{SU}(2) \to \mathrm{U}(\H_j)$ on $\H_j=\CC^{2j+1}$.
This is generated by hermitian operators $S_1,S_2,S_3$ on $\H_j$ with $[S_i,S_j]=\rmi \eps_{ijk}S_k$.
The full Lie algebra representation is of course $A\up j(X) = \sum_k \ip{X}{X_k} S_k$.
As usual, we denote the eigenbasis of $S_z$ by $\ket m$ with $m=-j,\ldots,j$ in unit steps. Here, $m$ is the eigenvalue, i.e.\ $S_z\ket m = m\ket m$.

Since the group $\mathrm{SU}(2)$ is compact, the Nelson Laplacian is the quadratic Casimir operator, see, for instance, \cite[Ch.~10]{Hall2015}.
By irreducibility (due to Schur's Lemma), it is a multiple of the identity, namely $\Delta \equiv \vec S^2 =j(j+1)\1$.
Therefore using the Nelson Laplacian, \cref{thm:main} gives the bound
\begin{equation}
    \norm{U\up j_g - U\up j_h} \leq j(j+1) \, \norm{\log(g^{-1}h)}_2.
\end{equation}
However, it is possible to find a tighter bound through \cref{thm:corollary}.
For any unit vector $X\in\mathfrak{su}(2)$, there is some $g\in\mathrm{SU}(2)$ such that $gXg^{-1} = X_3$.
Therefore, all $A\up j(X)$ are unitarily equivalent to $\norm{X}_2 S_z$.
In particular, we have $\norm{A\up j(X)} = j \norm{X}_2$.
Thus, by \cref{thm:corollary},
\begin{equation}
        \norm{U\up j_g - U\up j_h} \leq j^2\, \norm{\log(g^{-1}h)}_2.
\end{equation}
From the above discussion, it is evident that this bound is tight in first order. 
Notice that the relative difference between the bounds obtained from using the Nelson Laplacian $\Nelson$ and from using the alternative operator $K=j^2\1$ only grows linearly in $j$.

\subsection{Special orthogonal group and FLO representation}\label{sec:free_fermion}

Let us study the Lie group 
\begin{equation}
    \mathrm{SO}(2m)=\set{Q\in\RR^{2m\times2m}\given Q^\top Q=\1,\,\det(Q)=1}.
\end{equation}
We will need to look at the universal cover of $\mathrm{SO}(2m)$, which is known as the spin group and will be denoted by $\mathrm{Spin}(m)$.
Its Lie-algebra $\mathfrak{so}(2m)=\set{X\in\RR^{2m\times2m}\given X=-X^\top }$ is formed by real anti-symmetric $2m\times 2m$ matrices. 
An Ad-invariant inner product is given by the Frobenius inner product $\langle X,Y\rangle=\tr[X^\top Y]$. 
Note that $\mathrm{SO}(2)\cong \mathrm{U}(1)$, the circle group.

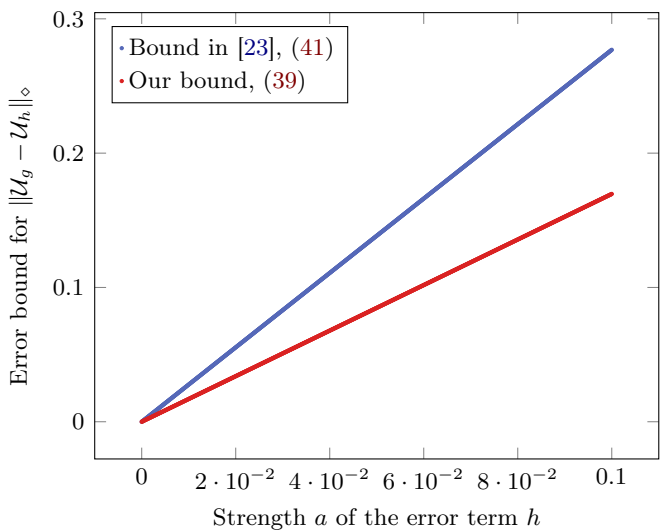
\begin{figure}
\centering
  	\begin{tikzpicture}[mark size={0.6}, scale=1]
  	\pgfplotsset{
    width=0.6\textwidth,
    height=0.5\textwidth
    }
	\begin{axis}[
	ylabel near ticks,
	xlabel={Strength $a$ of the error term $h$},
	ylabel={Error bound for $\Vert \U_g-\U_h\Vert_\diamond$},
	x post scale=1,
	y post scale=1,
	legend pos=north west,
	legend cell align={left},
	legend columns=1,
	label style={font=\footnotesize},
	tick label style={font=\footnotesize},
	]
	\addplot[color=1, only marks] table[x=a, y=dist, col sep=comma]{SO2n_FLO_Bound.csv};
	\addplot[color=2, only marks] table[x=a, y=dist, col sep=comma]{SO2n_Our_Bound_Log.csv};
	\legend{\footnotesize Bound in \cite{Oszmaniec2022}\text{,}   \eqref{eq:FLO}, \footnotesize Our bound\text{,} \eqref{eq:SO_channel_compare_log}};
	\end{axis}
	\end{tikzpicture}
  	\caption{Comparison of our bound for $\mathrm{SO}(2m)$ in \cref{eq:SO_channel_compare_log} (red) with the bound in \cite[Eq.~(F1)]{Oszmaniec2022} (blue) for two particular channels. In this numerical simulation, we consider $m=2$, $g=\exp(B_{21})$ and $h=\exp(B_{21}+aB_{31})$ with $B_{jk}$ defined in \cref{eq:so_basis}. We find that our bound is tighter (red vs.\ blue) in this case.}
  	\label{fig:SO_compare}
\end{figure}

In order to construct the free-fermion representation of $\mathrm{SO}(2m)$, consider the fermionic Fock space
\begin{equation}
	\H = \mathfrak{F}_-(\CC^m) \coloneqq \bigoplus_{j=0}^m \left(\CC^m\right)^{\wedge j},
\end{equation}
where $\wedge j$ denotes the $j$-fold anti-symmetric tensor product of $\CC^m$ and $\left(\CC^m\right)^{\wedge0}=\CC$. 
Note, that $\dim(\H) = 2^m$.
In particular, all appearing operators are bounded.
We denote the fermionic annihilation and creation operators by $b_1,\dots,b_m$ and $b_1^*,\dots,b_m^*$, respectively.
They satisfy the canonical anti-commutation-relations
\begin{equation}
	\set{b_j,b_k^*}=\delta_{jk}\1,\quad \set{b_j,b_k}=0.
\end{equation}
Here, $\set{X,Y}=XY+YX$ is the anti-commutator, and $\delta_{jk}$ is the Kronecker delta. From those operators, we define an equivalent set of $2m$ Majorana operators
\begin{align}
	c_{2j-1}=b^*_j+b_j
	\qandq
	c_{2j}=-\rmi(b_j^*-b_j).
\end{align}
A quick calculation shows that the inverse relations read
\begin{align}
	b_j=\frac{1}{2}(c_{2j-1}-\rmi c_{2j})
	\qandq
	b_j^*=\frac{1}{2}(c_{2j-1}+\rmi c_{2j}).
\end{align}
The Majorana operators satisfy the Clifford algebra relations
\begin{equation}
	\set{c_j,c_k}=2\delta_{jk}\1.\label{eq:Clifford}
\end{equation}
Now, let $c=(c_1,\dots,c_{2m})$ be the vector of all Majorana operators and define
\begin{equation}
	\dU(X)=\frac{\rmi}{4}\,c \cdot Xc\label{eq:Hamiltonian_free_fermion}
\end{equation}
for $X\in \mathfrak{so}(2m)$. One has $A(X)^*=A(X)$ and $[A(X),A(Y)] = \rmi A([X,Y])$. 
Therefore, $A$ is a faithful representation of $\mathfrak{so}(2m)$ by self-adjoint generators on $\H$.
The Lie algebra $-\rmi A(\mathfrak{so}(2m)) \subset \mathfrak{u}(\H)$ generates a connected subgroup of the unitary group $\mathrm{U}(\H)$, which is isomorphic to $\mathrm{Spin}(2m)$ \cite{Oszmaniec2022} and this isomorphism can be viewed as a representation
\begin{equation}\label{eq:FLO_rep}
    U : \mathrm{Spin}(2m) \to \mathrm{U}(\H).
\end{equation}
Since $\mathrm{Spin}(m)$ is a covering group of $\mathrm{SO}(2m)$, one can view \cref{eq:FLO_rep} as a projective representation of $\mathrm{SO}(2m)$ as explained in \cref{sec:projective}.
Here we will consider the induced unitary channel representation $(\U,\H)$ of $\mathrm{SO}(2m)$, i.e.\ $\U_g(\rho) = U_{\tilde g} \rho U_{\tilde g}^*$ for some $\tilde g \in \pi^{-1}(\set g)$ where $\pi : \mathrm{Spin}(2m)\to \mathrm{SO}(2m)$ is the covering map.
The Nelson Laplacian is (see \cref{appendix:Nelson_SO})
\begin{align}
	\Nelson=\frac{m(2m-1)}{8}\1,
\end{align}
 and we obtain
\begin{equation}
	\left\Vert\left(U_g-U_h\right)\psi\right\Vert\leq\sqrt{m(2m-1)/8}\,\norm\psi\ d(g,h),\label{eq:so_bound}
\end{equation}
for $g,h \in \mathrm{SO}(2m)$, which are close to each other (cf.\ \cref{thm:covering}).
If $\log(g^{-1}h)$ exists, $d(g,h)\leq \Vert\log(g^{-1}h)\Vert_2$, where $\norm B_2=\sqrt{\tr[B^\top B]}$ is the Frobenius norm.
By \cref{thm:riemannian_metric}, one even has $d(g,h) = \norm{\log(g^{-1}h)}_2$ if $g$ and $h$ are sufficiently close.
In any case, we have for the unitary channels associated to $g,h \in \mathrm{SO}(2m)$
\begin{equation}
	\left\Vert \U_g-\U_h\right\Vert_\diamond\leq\sqrt{m(2m-1)/2} \ d(g,h). \label{eq:SO_channels}
\end{equation}
Again, we have $d(g,h)\leq \Vert\log(g^{-1}h)\Vert_2$ if $\log(g^{-1}h)$ exists. In this case, \cref{eq:SO_channels} can be further manipulated by using the fact that for any orthogonal matrix $Q$, we have the relation \cite[Exercise~B.5]{Aubrun2017}
\begin{equation}\label{eq:log_Frobenius_bound}
	\norm{\log(Q)\Vert_2\leq\frac{\pi}{2}} Q-\1\Vert_2.
\end{equation}
Therefore, we find
\begin{align}
	\left\Vert \U_g-\U_h\right\Vert_\diamond&\leq \sqrt{m(2m-1)/2} \ \Vert\log(g^{-1}h)\Vert_2\label{eq:SO_channel_compare_log}\\
	&\leq \frac{\pi}{2\sqrt{2}}\sqrt{m(2m-1)} \ \Vert g-h\Vert_2,\label{eq:SO_channel_compare}
\end{align}
where we have used that $\Vert QB\Vert_2=\sqrt{\tr[ B^\top Q^\top Q B]}=\sqrt{\tr[B^\top B]}=\norm B_2$ holds for all $Q\in\mathrm{SO}(2m)$ and for all $B\in\RR^{2m\times2m}$.
A similar bound has been found in Ref.~\cite[Eq.~(F1)]{Oszmaniec2022}, which states that
\begin{equation}\label{eq:FLO}
	\norm{\U_g-\U_h}_\diamond \leq 2m\Vert g-h\Vert_\infty,
\end{equation}
where $\norm B_\infty=\sup_{\abs v =1}\abs{Bv}$ is the operator norm and $\abs\placeholder$ is the Euclidean length on $\RR^{2m}$. 
We compare the bound in Ref.~\cite[Eq.~(F1)]{Oszmaniec2022} with ours from \cref{eq:SO_channel_compare_log} numerically in \cref{fig:SO_compare}. Our bound (red) seems to be better than the result in Ref.~\cite[Eq.~(F1)]{Oszmaniec2022} (blue).

\subsection{Displacement operators}\label{sec:displacement}

For pedagogical reasons, we consider a quantum system of $m$ canonical degrees of freedom in the Schr\"odinger representation $\H=L^2(\RR^{m})$ and denote by $Q_i$ and $P_i$ the canonical position and momentum operators.
The Lie group $G$ in this example is the group of phase space translations, which is naturally isomorphic to $(\RR^{2m},+)$. Efficient bounds for this example have been found in \cite{Becker2021}.
For $\xi \in \RR^{2m}$ consider $D(\xi) = e^{-i\xi \cdot \Omega R}$, where $R = (Q_1, \dots, Q_m, P_1, \dots, P_m)$ and where $\Omega$ is the symplectic matrix 
\begin{equation}\label{eq:symplectic_matrix}
 \Omega=\begin{pmatrix} 0 & \1_m\\  -\1_m & 0 \end{pmatrix}.
\end{equation}
The $D(\xi)$ are called \emph{displacement operators} (or \emph{Weyl operators}) and implement phase space translations.
They form a strongly continuous projective unitary representation of the connected and simply connected group $G = \RR^{2m}$ with
\begin{align}\label{eq:weyl_rel}
D(\xi)D(\eta) = e^{-\tfrac \rmi 2 \xi\cdot\Omega\eta}D(\xi + \eta).
\end{align}
\cref{eq:weyl_rel} is known as the \emph{Weyl relations}. We write $\xi \cdot R = \sum_j \xi_j R_j$, which is simply the Lie algebra representation $A(\xi) = \xi\cdot R$.
The Weyl relations are an integrated version of the canonical commutation relations that can be expressed as $[\xi\cdot R,\eta\cdot R] = \rmi\xi\cdot \Omega\eta\, \1$.

Another way to view the Weyl relations is that the displacement operators are a (proper) unitary representation of the \emph{Heisenberg group} $\mathbb H_m$, which is the central extension of $\RR^{2m}$ corresponding to the projective representation discussed above.
The Heisenberg group is the set $\mathbb H_m = \RR^{2m+1} = \RR^{2m}\times\RR$ with the non-commutative group operation $$\mat{\xi\\ t}*\mat{\eta\\ s} = \mat{\xi+\eta\\ t+s-\frac12 \xi\cdot\Omega\eta}.$$
Its Lie algebra is the vector space $\RR^{2m+1}$ with the Lie bracket $[(\xi,x),(\eta,y)] = (0,\xi\cdot\Omega\eta)$, which we equip with the Euclidean inner product.
As we want to give a bound on the quantity $\norm{D(\xi)\psi-D(\eta)\psi}$, we are interested in the corresponding metric distance
\begin{equation}
    d\bigg(\mat{\xi\\0},\mat{\eta\\0}\bigg)=d\bigg(\mat{-\eta\\0}*\mat{\xi\\0},\mat{0\\0}\bigg) 
    = d\bigg(\mat{\xi-\eta\\ \tfrac12\eta\cdot\Omega\xi},\mat{0\\0}\bigg).
\end{equation}
From the fact that the Lie-theoretic exponential map is just the identity, we get that
\begin{equation}
	d\bigg(\mat{\xi\\0},\mat{\eta\\0}\bigg)=\sqrt{\abs{\xi-\eta}^2 + \tfrac14 (\xi\cdot\Omega\eta)^2}.
\end{equation}
In order to get a concrete bound, we also need to know the Nelson Laplacian. 
For this, we pick the standard basis of $\RR^{2m+1}$ and obtain $\Delta = \1+\sum_{i=1}^m (P_i^2 + Q_i^2)$, which is precisely the identity plus twice the quantum harmonic oscillator $H$.
It can, however, be readily checked that one may drop the ``$+\1$'' without violating \cref{eq:beta_ineq}. 
As explained in \cref{thm:corollary}, we may thus use $\N=2H = \Nelson -\1$ instead of the Nelson Laplacian.
Inserting this into \cref{eq:main_bound} gives the following estimate
\begin{equation}
    \norm{D(\xi)\psi-D(\eta)\psi }^2 \le 2\,\big(\abs{\xi-\eta}^2 + \tfrac14 (\xi\cdot\Omega\eta )^2 \big)\, \ip\psi{H\psi},
\end{equation}
valid for all $\psi\in\H$.

\subsection{Quasi-free bosonic transformations}\label{sec:metaplectic}

We again consider a quantum system of $m$ canonical degrees of freedom but this time the symmetry group is the symplectic group 
\begin{equation}
    \mathrm{Sp}(2m,\RR)=\set{M\in\RR^{2m\times 2m} \given M^\top \Omega M=\Omega},
\end{equation}
where $\Omega$ denotes the symplectic matrix as in \cref{eq:symplectic_matrix}.
The symplectic group is a non-compact connected simple Lie group. 
An important role in this example is played by the double cover of the symplectic group, known as the metaplectic group $\textrm{Mp}(2m,\RR)$~\cite{Li2000}\@.
We denote the covering homomorphism by $\pi: \textrm{Mp}(2m,\RR) \to \textrm{Sp}(2m,\RR)$.
The natural representation of the symplectic group in quantum mechanics is a projective representation, which comes from a proper representation of the metaplectic group.
It is widely used in quantum optics, where it corresponds to passive and active bosonic transformations.
The Lie algebra of the symplectic group -- and hence also of the metaplectic group -- is
\begin{equation}
    \mathfrak{sp}(2m,\RR)=\set{X\in\RR^{2m\times2m}\given \Omega X+X^\top \Omega=0}.
\end{equation}
We equip $\mathfrak{sp}(2m,\RR)$ with the Frobenius inner product $\langle X,Y\rangle=\tr[X^\top Y]$, which is not $\Ad$-invariant. In fact, there is no $\Ad$-invariant inner product on $\mathfrak{sp}(2m,\RR)$ \cite{Milnor1976}. 
As a vector space, $\mathfrak{sp}(2m,\RR)$ is isomorphic to the space of symmetric matrices $\mathrm{Sym}(2m,\RR)=\set{X\in\\\RR^{2m\times2m}\given X=X^\top }$ via the map $X \mapsto S=\Omega X$ and its inverse $S\mapsto X =-\Omega S$.
In fact, this isomorphism is isometric w.r.t.\ the Frobenius inner product.

For quantum optics, the subgroup $\mathrm{USp}(2m,\RR)\subset \mathrm{Sp}(2m,\RR)$ plays an important role. It is isomorphic to the unitary group $\mathrm{U}(m)$ under the standard identification $\RR^{2m}\cong \CC^m$ and corresponds exactly to the passive bosonic transformations.
The generators of $\mathrm{USp}(2m,\RR)$ are the skew-symmetric elements $X\in\mathfrak{sp}(2m,\RR)$ \cite[Ch.~4]{Folland2015}. Generators $X\in\mathfrak{sp}(2m,\RR)$ that are not skew-symmetric -- such as e.g.\ $X= \mathrm{diag}(1,-1)$ for $m=1$ -- correspond to active bosonic transformations, i.e.\ squeezing.

We are going to look at the \emph{metaplectic representation}, which is a strongly continuous projective representation of $\Sp(2m,\RR)$ on $\H = L^2(\RR^m)$.
We will continue to use the notation introduced in \cref{sec:displacement}, e.g.\ the canonical operators are denoted $R=(Q_1,\dots,Q_m,\\P_1,\dots,P_m)$.
Furthermore, let
\begin{equation}
	\dU(X)=\frac{1}{2}R \cdot \Omega XR\label{eq:dU_symplectic}
\end{equation}
with $X\in\mathfrak{sp}(2m,\RR)$.\footnote{In some parts of the literature, \cref{eq:dU_symplectic} is written in terms of bosonic creation and annihilation operators instead of canonical position and momentum operators. This is also known as the as the \emph{operator representation} of $\mathfrak{sp}(2m,\RR)$ \cite[Ch.~6.1]{Gilmore2008}\@.}
A common invariant domain for all $A(X)$ is given by the Schwartz functions $\mathscr{S}(\RR^m) \subset \H$~\cite[Ch.~V.3]{reed2012methods}\@.

It is straightforward to see that one has $[A(X),A(Y)] =  \rmi A([X,Y])$ on the Schwartz space $\mathscr S(\RR^m)$.
Exponentiation of $A(\mathfrak{sp}(2m,\RR))$ yields a subgroup of the unitary group on $\H$, which happens to be isomorphic to the metaplectic group. This isomorphism, which is a homeomorphism with respect to the strong topology, is known as the \emph{metaplectic representation}~\cite[Ch.~4]{Folland2015}\@.
We denote it by
\begin{equation}
    U : \mathrm{Mp}(2m,\RR) \to \mathrm{U}(\H).
\end{equation}

The double cover $\pi:\textrm{Mp}(2m,\RR) \to \textrm{Sp}(2m,\RR)$ does not admit a global continuous section, i.e.\ there is no continuous map $s:\mathrm{Sp}(2m,\RR) \to \mathrm{Mp}(2m,\RR)$ such that $\pi\circ s =\id$.
It is, however, possible to find a section $s$ which is continuous near the identity.
The (arbitrary) choice of such a section turns the representation $U$ into a projective unitary representation of the symplectic group, which is strongly continuous near the identity.

An easy example illustrating the fact that of \cref{eq:dU_symplectic} does not define a proper representation of the symplectic group is the following:
For $m=1$ consider the generator $h=-\Omega\in\mathfrak{sp}(2,\RR)$ of the subgroup
\begin{equation}
    \rme^{\alpha h} =\mat{\cos(\alpha) & -\sin(\alpha)\\\sin(\alpha)&\cos(\alpha)}.
\end{equation}
It is clearly the case that $\rme^{2\pi h} = \1_2\in\Sp(2,\RR)$.
However, the Hamiltonian assigned to $h$ is the harmonic oscillator
\begin{equation}
    A(h) = -\frac12 R\cdot\Omega^2 R = \frac12(Q^2+P^2)
\end{equation}
which exponentiates to $\rme^{-2\pi\rmi A(h)} =-\1$.
This is easily seen from the fact that the spectrum of the harmonic oscillator is $\frac12+\NN_0$ because $\rme^{-\rmi 2\pi (n+1/2)} =-1$.
Therefore, the Lie algebra representation $A$ cannot give rise to a proper unitary representation of the symplectic group but only to a projective representation.

\begin{figure}
 \centering
  	\begin{tikzpicture}[mark size={0.6}, scale=1]
  	\pgfplotsset{
    width=0.6\textwidth,
    height=0.5\textwidth
    }
	\begin{axis}[
	ylabel near ticks,
	xlabel={Trotter steps $L$},
	ylabel={Trotter error bound $\Vert \U_g-\U_h\Vert_\diamond^{\numop ,E}$},
	xmode = log,
	ymode = log,
	x post scale=1,
	y post scale=1,
	legend pos=north east,
	legend cell align={left},
	legend columns=1,
	label style={font=\footnotesize},
	tick label style={font=\footnotesize},
	]
	\addplot[color=1, only marks] table[x=L, y=dist, col sep=comma]{Symplectic_Becker_Bound.csv};
	\addplot[color=2, only marks] table[x=L, y=dist, col sep=comma]{Symplectic_Our_Bound.csv};
	\legend{\footnotesize Bound in \cite{Becker2021}\text{,}   \eqref{eq:bound_symplectic_Becker}, \footnotesize Our bound\text{,} \eqref{eq:symplectic_channel_compare}};
	\end{axis}
	\end{tikzpicture}
  	\caption{Comparison of our bound for the symplectic group in \cref{eq:symplectic_channel_compare} with the bound in \cite[Thm.~S12]{Becker2021} for a particular Trotter problem on a $\log$-$\log$ axis. 
  	That is, $g=\rme^{-t(X+Y)}$ and  $h=(\rme^{-tX/L} \rme^{-Yt/L})^L$ for $t\in\RR$ and $X,Y\in\mathfrak{sp}(2m,\RR)$. 
  	In this particular simulation, we chose $m=1$, $t=1$, $X=\Omega$ (harmonic oscillator) and $Y=\Omega\sigma_x$ (generator of the squeezing transformation), where $\sigma_x$ denotes the first Pauli matrix.
  	We compute the bounds for the energy-constrained diamond norm of the associated unitary channels up to the second Fock state, i.e., $\langle \psi,\numop \psi\rangle\leq 2$. The energy is taken with respect to the number operator $\numop $. Our bound correctly covers the $\mathcal{O}(1/L)$ scaling of the Trotter product formula.}
  	\label{fig:symplectic_compare}
\end{figure}
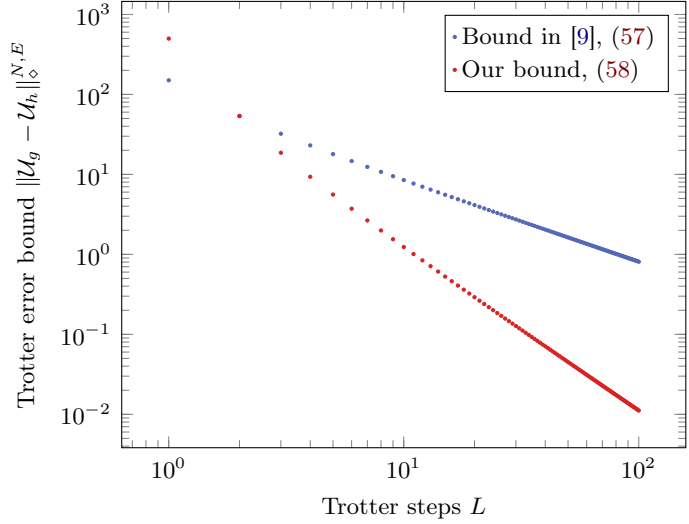

The Nelson Laplacian of the metaplectic representations is
\begin{equation}
	\Nelson = H^2+\frac{3m}{8}\1,
\end{equation}
where $H=\frac{1}{2}\sum_{k=1}^m(Q_k^2+P_k^2)$ denotes the $m$-mode harmonic oscillator. 
See \cref{appendix:Nelson_symplectic} for the derivation. Therefore, for $g,h\in \textrm{Sp}(2m, \RR)$, which are close enough to each other, we obtain (cf.\ \cref{thm:covering}) the bound
\begin{equation}
	\Vert (U_g-U_h)\psi\Vert\leq \sqrt{\ip\psi{H^2\psi}+\frac{3m}{8}}\  d(g,h),\label{eq:bound_symplectic}
\end{equation}
where $\norm\psi = 1$.
Consider $\psi$ to be a Fock state $\ket{\pmb{n}}$ with $\pmb{n} \in\NN_0^m$ being a multi-index. That is, $\ket{\pmb{n}}$ is an eigenstate of the number operator $\numop  =H-\frac{m}{2}\1$ according to the eigenvalue equation $\numop \ket{\pmb{n}}=\vert\pmb{n}\vert\,\ket{\pmb{n}}$. Then \cref{eq:bound_symplectic} becomes
\begin{equation}
	\Vert (U_g-U_h)\ket{\pmb{n}}\Vert\leq\sqrt{\left(\vert\pmb{n}\vert+\frac{m}{2}\right)^2+\frac{3m}{8}}\, d(g,h).
\end{equation}
 For the energy-constrained diamond norm, we obtain
 \begin{equation}
 	\Vert \U_g-\U_h\Vert_\diamond^{H^2,E}\leq 2\sqrt{E +\frac{3m}{8}}\,  d(g,h).\label{eq:bound_symplectic_channel}
 \end{equation}
 A similar bound is presented in the supplementary material of Ref.~\cite[Thm.~S12]{Becker2021},
 \begin{align}
 	\Vert \U_g-\U_h\Vert_\diamond^{\numop ,E}\leq &\ 2\sqrt{(\sqrt{6}+\sqrt{10}+5\sqrt{2}m)(E+1)}\nonumber \times \cdots\\
 	&\ \ \cdots \times \left(\sqrt{\frac{\pi}{\Vert g^{-1}h\Vert_\infty+1}}+\sqrt{2\Vert g^{-1}h\Vert_\infty}\right)\sqrt{\Vert g^{-1}h-1\Vert_2}.\label{eq:bound_symplectic_Becker}
 \end{align}
Here, $\Vert A\Vert_\infty=\sup_{\Vert\psi\Vert=1}\Vert A\psi\Vert$ is the operator norm while $\Vert A\Vert_2=\sqrt{\tr A^*A}$ is the Frobenious norm.
Let us compare this bound with \cref{eq:bound_symplectic_channel}. By following the steps in the proof of Ref.~\cite[Thm.~S12]{Becker2021}, we can further bound \cref{eq:bound_symplectic_channel} by
\begin{equation}
	\Vert \U_g-\U_h\Vert_\diamond^{H^2,E}\leq 2\sqrt{E+\frac{3m}{8}}\left(\frac{\pi}{\Vert g^{-1}h\Vert_\infty+1}+2\Vert g^{-1}h\Vert_\infty\right)\Vert g^{-1}h-1\Vert_2.\label{eq:symplectic_channel_compare}
\end{equation}
Both bounds have a term that depends only on the energy (first term), which is multiplied with a term measuring the distance between $g^{-1}h$ and the identity.
Notice, that our bound in \cref{eq:symplectic_channel_compare} involves $\Vert g^{-1}h\Vert_\infty$ and $\Vert g^{-1}h-1\Vert_2$ in the second term, whereas the bound from \cite[Thm.~S12]{Becker2021} depends on the square root of these quantities. 
In most cases, one is particularly interested in the tightness of the second term. As an example where the second term plays a crucial role, let us consider the study of Trotter errors \cite{Childs2021}.
In this context, $g=\rme^{-t(X+Y)}$ with $t\in\RR$ and $X,Y\in\mathfrak{sp}(2m,\RR)$ is some target dynamics, which should be approximated through the Trotter product formula given by $h=(\rme^{-\Omega Xt/L}\rme^{-\Omega Yt/L})^L$. 
Here, $\Vert g^{-1}h-1\Vert_2$ determines the asymptotic scaling of the Trotter product formula with the number of Trotter steps $L$. 
It is well-known that the Trotter error asymptotically scales as $\mathcal{O}(1/L)$ \cite{Childs2021,Suzuki1985}. This scaling is correctly captured by our bound in \cref{eq:symplectic_channel_compare}, whereas the bound in \cite[Thm.~S12]{Becker2021} leads to a $\mathcal{O}(1/\sqrt{L})$ behavior. 
We compare the two bounds explicitly for an example of such a Trotter problem in \cref{fig:symplectic_compare}.

\subsection{Two-mode bosonic representation of $\SU(1,1)$}\label{sec:su11}

The idea of $\SU(1,1)$ interferometry was first proposed in \cite{su11interferometers} and has been a major topic in quantum optics since then.
By incorporating squeezing, $\SU(1,1)$ interferometry acquires significant advantages over conventional interferometry \cite{ou2020quantum}.
Here conventional interferometry means a Mach-Zehnder-type setup with beam splitters and relative phase shifts. In such a setup, the implementable transformations are parameterized by $\SU(2)$, and one sometimes refers to this setup as $\SU(2)$-interferometry \cite{su11interferometers}.
The setup of $\SU(1,1)$ interferometry is similar: one replaces the two beam splitters of the Mach Zehnder interferometer by two parametric amplifiers \cite{ou2020quantum}.
In this case, the implementable transformations correspond to elements of $\SU(1,1)$.
The Hilbert space for both of these setups is the two-mode bosonic Fock space $\H =\mathcal F_+(\CC^2)$.

Of course, all transformations that can be implemented in both of these interferometry setups are just special cases of the two-mode quasi-free bosonic transformations for which we obtained bounds in \cref{sec:metaplectic}, but the error estimates that we obtain in this way are unnecessarily large.
The reason for this is that the Nelson Laplacian of the metaplectic representation is built so that \cref{eq:bertaN} holds for all $X\in\mathfrak{sp}(4,\RR)$ even though we only need this assumption on a Lie subalgebra, namely $\su(1,1)$.

To avoid confusion, we briefly recall the definition of the Lie group $\SU(1,1)$: It is the matrix group of complex $2\times2$-matrices of the form
\begin{equation}
    \mat{\alpha & \beta \\\Bar\beta&\Bar\alpha } \quad\text{with}\ \abs\alpha^2 - \abs\beta^2=1.
\end{equation}
Its Lie algebra $\mathfrak{su}(1,1)$ consists of trace-less matrices $X$ such that $X \sigma_z + \sigma_z X^\dagger =0$, where $\sigma_z = \mathrm{diag}(1,-1)$ is the third Pauli matrix. A basis of $\mathfrak{su}(1,1)$ is
\begin{equation}
    X_0 = \frac \rmi2\mat{ 1&0\\0&-1},\quad X_1 =  \frac\rmi2 \mat{0&-1\\1&0} \qandq X_2 =\frac12\mat{0&1\\1&0}.
\end{equation}
We equip the Lie algebra with the Hilbert-Schmidt inner product $\ip XY = 2\tr[X^\dagger Y]$ for $X,Y\in\su(1,1)$ so that $\set{X_0,X_1,X_2}$ indeed is an orthonormal basis.
The norm on the Lie algebra is thus $\norm{X}_{\su(1,1)} = \sqrt2 \norm X_2 = \sqrt2 \sqrt{\tr[X^*X]}$.
The $X_i$ satisfy the commutation relations 
\begin{equation}
    [X_0,X_1] = X_2,\quad [X_0,X_2] = -X_1 \qandq [X_1,X_2] = - X_0.
\end{equation}
It is a mathematical fact that the Lie groups $\SU(1,1)$ and $\mathrm{Sp}(2,\RR)$ (and $\mathrm{SL}(2,\RR)$ for that matter) are isomorphic.
We can, however, regard this as a mathematical curiosity.

To define the representation we consider the two-mode bosonic Fock space $\H = \mathcal F_+(\CC^2)$.
We denote the creation, annihilation and number operators by $a_i^\dagger$, $a_i$ and $N_i$, $i=1,2$, respectively.
Consider the difference of number operators
\begin{equation}
    D = N_1-N_2.
\end{equation}
The bosonic Fock space decomposes into a direct sum of its eigenspaces
\begin{equation}
    \H = \bigoplus_{n\in\ZZ}\mathcal D_n \quad\text{with}\ \mathcal D_n = \Bar{\mathrm{span}} \set[\big]{\,\ket{n_1,n_2} \given n_1,n_2\in\NN_0\ \text{s.t.}\ n_1-n_2=n},
\end{equation}
where $\ket{n_1,n_2}$ denote the two-mode Fock states.
Consider the following bosonic operators
\begin{equation}
    K_0 = \frac12 (N_1+N_2 +\1), \quad
    K_1 = \frac12(a_1^\dagger a_2^\dagger + a_1 a_2) \qandq
    K_2 = -\frac \rmi2 (a_1^\dagger a_2^\dagger - a_1 a_2).
\end{equation}
Note that $K_0$ is just the two-mode Harmonic oscillator.
The map $A : X \mapsto \sum_i \ip{X}{X_i}K_i$ is a representation of $\mathfrak{su}(1,1)$ since 
\begin{equation}
    [K_0,K_1] = \rmi K_2,\quad [K_0,K_2] = -\rmi K_1 \qandq [K_1,K_2] = - \rmi K_0.
\end{equation}
Each of the $K_i$ commutes with $D$ so that we may restrict them to the subspaces $\mathcal D_n$.
These restrictions exponentiate to strongly continuous irreducible representations $(U\up n,\mathcal D_n)$ of $\mathrm{SU}(1,1)$ which are labelled by $n \in\ZZ$ \cite{lieb2019wehrl}.
It follows from the canonical commutation relations that $K_1^2+K_2^2 = K_0^2 - \frac14 D^2 +\frac14\1$.
For the full bosonic representation $(U,\H) = \bigoplus_{n\in\ZZ}(U\up n , \mathcal D_n)$, we get $\Nelson = 2K_0^2 - \tfrac14( D^2 -\1)$.
Combining this with $D=n$ on $\mathcal D_n$,  proves that the Nelson Laplacian $\Nelson\up n$ of $(U\up n,\mathcal D_n)$ takes the form
\begin{align}
    \Nelson\up n &= K_0^2 + K_1^2 + K_2^2 = 2K_0^2 + \frac{1-n^2}4\,\1.
\end{align}
Note that the operators only act on $\mathcal D_n$, which actually allows the above equation to define a positive operator for all $n\in\ZZ$.
\cref{thm:main} applies and shows that
\begin{equation}
    \norm{U\up n_g \psi - U\up n_h \psi } \leq  \frac{1}{2}\sqrt{ 8 \ip\psi{K_0^2 \psi} -n^2 + 1} \ d(g,h)
\end{equation}
Recall that $K_0$ is just the two-mode Harmonic oscillator restricted to $\mathcal D_n$.
Note that one can always use $d(g,h) \leq \sqrt{2}\norm{\log(g^{-1}h)}_2$ for sufficiently close $g,h$ to obtain practical bounds.
Consider, for example, a Fock state $\psi = \ket{n_1,n_2}$. Then we get the following
\begin{equation}
    \norm{U_g \ket{n_1,n_2} - U_h \ket{n_1,n_2} }
    \leq \frac{1}{\sqrt2}\sqrt{8(n_1+n_2+1)^2-(n_1-n_2)^2+1} \ \norm{\log(g^{-1}h)}_2.
\end{equation}
For the corresponding unitary channels, one obtains the following bounds on the energy-constrained diamond norm with respect to $K_0^2$, the square of the harmonic oscillator,
\begin{equation}
    \norm{\U\up n_g-\U\up n_h}_\diamond^{K_0^2,E} \leq \sqrt{8 E - n^2+1} \ d(g,h). 
\end{equation}

\subsection{Representation of the Lorentz group}\label{sec:Lorentz}

\renewcommand{\pmb}[1]{#1}
Consider the Minowksi space $\RR^{1,3}$ equipped with the scalar product
\begin{equation}
    (p,q) \mapsto \eta(p,q) = \vec p \cdot \vec q - p_0 q_0,
\end{equation}
where $\vec p = (p_1,p_1,p_2)$ are the entries of a four vector $p = (p_0,p_1,p_2,p_3)\in\RR^{1,3}$.
We \emph{do not} use the Einstein summation convention, and we only use lower indices to avoid confusion as we have to use both Euclidean and Lorentzian geometry.
The scalar product corresponds to the matrix $\mathrm{diag}(-1,1,1,1)$ which we also denote by $\eta$, i.e.\ $\eta(p,q)=p\cdot \eta q$, where we use ``$\,\cdot\,$'' to denote the Euclidean inner product.
The Lorentz group $G=\mathrm O(1,3)$ is the set of (real) $4\times 4$ matrices $\Lambda \in \RR^{4\times4}$, which leave the scalar product invariant, i.e.\ an invertible matrix $\Lambda$ is in $\mathrm O(1,3)$ if and only if
\begin{equation}
	\Lambda^\top \eta\Lambda=\eta.
\end{equation}
We restrict our attention here to the connected component, $G = \SO^+(1,3)$, called the proper orthochronous Lorentz group (or the restricted Lorentz group). 
The Lie algebra $\lie = \so(1,3)$ consists of trace-less matrices $X\in\RR^{4\times4}$ such that $X^\top \eta +\eta X=0$.
We equip it with half the Frobenius inner product $\ip XY = \frac12 \tr[X^\top Y]$, so that the norm becomes $\norm{\placeholder}_{\so(1,3)}=2^{-1/2}\norm{\placeholder}_2$.

The strongly continuous irreducible representations of the Lorentz group have been classified by Wigner \cite{Wigner1939}. 
Up to unitary equivalence, they depend on a ``mass'' $m\ge0$ and a ``spin'' $s\in\frac12\NN_0$ \cite[Ch.~I.3]{Haag2012}. Here, we consider the scalar, i.e.\ $s=0$, case only.
Error bounds for the representations with spin can be obtained similarly.
Consider the positive-energy (or ``future directed'') mass hyperboloid $\M_m=\set{\pmb{p}\in\RR^{1,3}\given \eta(p,p) =-m^2,\, p_0\geq 0}$ corresponding to mass $m>0$. 
Equivalently, $\M_m$ consists of those $\pmb p$ such that $p_0=\sqrt{\vec p^{\, 2}+m^2}$ and this gives us a coordinatization of $\M_m$ in terms of $\RR^3$.
Hence, we may view $p_0$ as a function of $\vec p$ for $p\in\M_m$.
In momentum space, the scalar representation of the Lorentz group acts on the Hilbert space $\H$ of square-integrable functions $\psi:\M_m\to \CC$, i.e.\ $\H=L^2(\M_m)$ with respect to the measure $\mathrm{d}^3\pmb{p}/(2p_0)$.
A Lorentz transformations $\Lambda$ acts on a wavefunction $\psi\in \H$ in the natural way
\begin{equation}
	U_\Lambda\psi(\pmb{p})=\psi(\Lambda^{-1} \pmb{p}).
\end{equation}
The irreducibility of the representation expresses itself by the Fourier transformed wave functions $\tilde\psi$ in position space satisfying the Klein-Gordon equation $( \partial_t^2-\Delta)\tilde\psi = -m^2\tilde\psi$.
We now compute the induced representation of the Lie algebra
\begin{equation}
	A(X)\psi(\pmb p) = \rmi \frac{\mathrm{d}}{\mathrm{d}t}\psi(\rme^{-tX}\pmb{p})\big|_{t=0} =-\rmi\nabla_{X\pmb p}\psi(\pmb{p}) ,\quad X\in\mathfrak{so}(1,3),
\end{equation}
where $\nabla_v$ denotes the directional derivative (in a direction $v$ tangent to $\M_m$) and $\psi$ is suitably differentiable. 
Indeed, $X\pmb p$ is always tangent to the (positive-energy) mass hyperboloid $\M_m$ at the point $p\in \M_m$ if $X\in \so(1,3)$.

To state our bounds, we need to compute the expectation values of the Nelson Laplacian for arbitrary $\psi\in\H$.
We will do this directly without first computing the action of the Nelson Laplacian on wavefunctions.
To this end, we will need an orthonormal basis $\set{X\up n}_n$ of $\so(1,3)$.
To obtain a useful expression for the expectation values $\ip\psi{\Nelson\psi}$ for differentiable $\psi$, we do the following:
We extend the wavefunction $\psi$ to a differentiable function (also denoted by $\psi$) on the full Minkowski space, and we extend the generators $A(X)$ to first-order differential operators on the full Minkowski space by setting $A(X)\psi(p) = -\rmi\sum_{ij} X_{ij}p_j\partial_i\psi$.
Subject to the condition that $\abs{\nabla\psi}$ should still be square-integrable on $\M_m$, this extension may be chosen arbitrarily.
We may now use the flat coordinates of the Minkowski space for our computation instead of having to compute derivatives in the coordinates of the curved 3-manifold $\M_m$.
Since we will integrate only over the sub-manifold $\M_m$ to compute the expectation value, it is guaranteed that the result will be independent of the arbitrary choice we made when we extended $\psi$.
We define the shorthand notation $T_{ijkl} = \sum_n X_{ij}\up n X_{kl}\up n$.
With this, we compute
\begin{align}\label{eq:NL_Lorentz_expectation}
    \ip\psi{\Delta\psi} &= \sum_n \ip{A(X\up n)\psi}{A(X\up n)\psi}\nonumber\\
    &= -\sum_{ijkln} \ip{X_{ij}\up n p_j \partial_i \psi}{X_{kl}\up n p_l\partial_k\psi}\nonumber\\
    &= -\sum_{ijkl} T_{ijkl} \ip{\partial_i\psi}{p_jp_l\partial_k\psi}.
\end{align}
Note that the partial derivatives $\partial_i\psi$ are in $\H$ because the extension is such that $\abs{\nabla\psi}$ is square-integrable on $\M_m$.
To compute further, we need to compute the numbers $T_{ijkl}$. To do this, we choose the standard basis of $\so(1,3)$:
\begin{equation}
\begin{aligned}
	J_1&=\begin{pmatrix}
		0 & 0 & 0 & 0\\
		0 & 0 & 0 & 0\\
		0 & 0 & 0 & -1\\
		0 & 0 & 1 & 0
	\end{pmatrix},\\[5pt]
	J_2&=\begin{pmatrix}
		0 & 0 & 0 & 0\\
		0 & 0 & 0 & 1\\
		0 & 0 & 0 & 0\\
		0 & -1 & 0 & 0
	\end{pmatrix},\\[5pt]
	J_3&=\begin{pmatrix}
		0 & 0 & 0 & 0\\
		0 & 0 & -1 & 0\\
		0 & 1 & 0 & 0\\
		0 & 0 & 0 & 0
	\end{pmatrix},
	\end{aligned}
	\quad\quad\quad
	\begin{aligned}
	K_1&=\begin{pmatrix}
		0 & 1 & 0 & 0\\
		1 & 0 & 0 & 0\\
		0 & 0 & 0 & 0\\
		0 & 0 & 0 & 0
	\end{pmatrix},\\[5pt]
	K_2&=\begin{pmatrix}
		0 & 0 & 1 & 0\\
		0 & 0 & 0 & 0\\
		1 & 0 & 0 & 0\\
		0 & 0 & 0 & 0
	\end{pmatrix},\\[5pt]
	K_3&=\begin{pmatrix}
		0 & 0 & 0 & 1\\
		0 & 0 & 0 & 0\\
		0 & 0 & 0 & 0\\
		1 & 0 & 0 & 0
	\end{pmatrix}.
	\end{aligned}
\end{equation}
The $J_i$ generate the subgroup $\SO(3)$ of spatial rotations while the $K_i$ generate boosts.
We use the standard notation $M_{ij}=\sum_k \eps_{ijk}J_k$ and $M_{0i}=-M_{i0} = K_i$ for $i,j,k\in\set{1,2,3}$.
These matrices satisfy $M_{mn}=-M_{nm}$ with $m,n=0,\ldots,3$, and we have the following formula for their matrix entries
\begin{equation}
    (M_{mn})_{ij} = \eta_{ni}\delta_{mj} - \eta_{mi}\delta_{nj}. 
\end{equation}
This gives
\begin{align}
    T_{ijkl} &= \frac12 \sum_{mn}(M_{mn})_{ij} (M_{mn})_{kl} \nonumber\\&
    = \frac12 \sum_{mn} ( \eta_{mi}\delta_{nj} - \eta_{ni}\delta_{mj})( \eta_{mk}\delta_{nl} - \eta_{nk}\delta_{ml}) \nonumber\\&
    =\frac12 \Big(\sum_m\eta_{mi}\eta_{mk}\delta_{jl} -\eta_{li}\eta_{jk}-\eta_{li}\eta_{jk} + \sum_n\eta_{ni}\eta_{nk} \delta_{jl}\Big) \nonumber\\&
    =\delta_{ik}\delta_{jl} -\eta_{li}\eta_{jk}.
\end{align}
Inserting this into \eqref{eq:NL_Lorentz_expectation} leads to
\begin{align}
    \ip\psi{\Delta\psi} 
    &=-\sum_{ijkl} T_{ijkl} \ip{\partial_i\psi}{p_jp_l\partial_k\psi} \nonumber\\
    &=-\sum_{ij} \ip[\big]{\partial_i \psi}{p_j^2 \partial_i\psi} + \ip[\Big]{\sum_{ij} p_j\eta_{ij}\partial_i\psi}{\sum_{kl}p_l\eta_{kl}\partial_k\psi} \nonumber\\
    &= \norm{\eta(p,\nabla\psi)}^2 - \norm{\abs p \nabla\psi}^2,\label{eq:NL_Lorentz_expectation2}
\end{align}
where we restricted $\psi$ again to $\H$. Notice that $\nabla\psi$ in the second term is a $\CC^4$-valued function. 
Thus, the second norm is taken in $L^2(\M_m,\CC^4)=\H\ox\CC^4$.
We stress again that even though \cref{eq:NL_Lorentz_expectation2} only makes sense if $\psi$ is extended to a differentiable function on $\RR^{1,3}$ (or at least some open subset containing $\M_m$), the value is independent of this extension.
By \cref{thm:main}, we obtain the following error bounds for Lorentz transformations in the scalar representation 
\begin{align}
	\Vert (U_\Lambda-U_{\tilde{\Lambda}})\psi\Vert 
	& \leq \frac1{\sqrt2}\sqrt{\norm{\eta(p,\nabla\psi)}^2 - \norm{\abs p \nabla\psi}^2}\ \norm{\log(\tilde\Lambda^{-1}\Lambda)}_2.
\end{align}

\section{Proofs}\label{sec:proofs}

In this section, we present the proofs of all the previous results. We will do this in the chronological order of how they appear in the preceding sections. Notice, however, that this order differs from the logical one as we might use later results in order to prove earlier ones. The logical order of the proofs is \cref{thm:lemma_nelson}, \cref{thm:main}, \cref{thm:lemma_metric}, \cref{thm:riemannian_metric}, \cref{lem:dim2}, \cref{thm:channel_bound}, \cref{thm:covering}.
Also, we will not explicitly prove \cref{thm:corollary} since the proof coincides with that of \cref{thm:main}.

\begin{proof}[Proof of \cref{thm:lemma_metric}]
    It is straightforward to see that $d$ is left-invariant, positive, symmetric (in the sense that $d(g,h)=d(h,g)$) and that $d$ satisfies the triangle inequality. 
    Consider the left-regular representation $(U,\H)$ which is defined on $\H=L^2(G,\mathrm d\mu)$, where $\mu$ is the left Haar measure of $G$, via $U_g\psi(h) = \psi(g^{-1}h)$. 
    It is well-known that the left-regular representation is faithful and strongly continuous.
    Strong continuity follows, for example, from \cite[Lem.~3.3.7]{ReiterStegeman2000} (faithfulness also follows from what is shown below).
    Suppose that $d(g,h)=0$, then our bound \cref{eq:main_bound} implies that $U_g=U_h$ in the left-regular representation, which by faithfulness yields $g=h$. 
    This concludes the proof for $d$ being a metric.
    
    To show that the metric topology induced by $d$ is the Lie group topology of $G$, we have to show that $g_n\to 1$ is equivalent to $d(g_n,1)\to0$ for any sequence $g_n$ in $G$. 
    If $g_n\to1$, then for large enough $n$ we have $\log(g_n)\to 0$ in $\lie$. 
    One also has the estimate $\norm{\log g_n}_\lie \geq d(g_n,1)$ since $g_n= e^{\log g_n}$ for large enough $n$. Combining these two shows that $g_n\to1$ implies $d(g_n,1)\to0$.

    For the converse, consider again the left regular representation. 
    We will prove the contraposition, i.e.\ that if $g_n$ is a sequence that does not converge to $1$, then also $U_{g_n}$ does not converge to $\1$.
    Due to \cref{eq:main_bound}, this will show that $d(g_n, 1)$ does not converge to $0$.
    To this end, we first observe that if $V$ is a neighborhood of $1$, then there exists a neighborhood $W\subset V$ of $1$ such that for each $g\not\in V$ it holds that $gW\cap W = \varnothing$.
    This follows from existence of a symmetric neighborhood $W\subset V$ of $1$ such that $W^2 = \set*{gh\given g,h\in W}\subset V$ \cite{ReiterStegeman2000}:
    If $gW\cap W\neq \varnothing$ then there exists a $h\in W$ such that $gh\in W$.
    But since $W$ is symmetric, it then follows that $g = (gh)h^{-1}\in W^2\subset V$.
    Thus $g\not\in V$ implies $gW\cap W = \varnothing$.
    Let $\psi = \chi_W$ be the characteristic function of $W$.
    By the construction of $W$, we then have that $\norm{U_{g}\psi - \psi}_2^2 = 2\norm{\psi}^2=2\mu(W) > 0$ for each $g\not\in V$, where $\mu$ is the Haar measure on $G$.
    In particular, since for each $1\neq g\in G$, there exists a neighborhood $V$ of $1$ that does not contain $g$, faithfulness follows.
    Now, let $g_n$ be a sequence in $G$, not converging to $1$.
    By passing to a subsequence we can assume that there exists a neighborhood $V$ of $1\in G$ such that $g_n\not\in V$ for all $n$.
    From what we have seen before, it follows that $U_{g_n}$ does not converge strongly to $\1$.
\end{proof}

\begin{proof}[Proof of \cref{thm:riemannian_metric}]
    It is known that exponential curves are geodesics with respect to a bi-invariant metric and that conversely, any geodesic is of the form $t\mapsto g \rme^{tX}$ for some $X\in\lie$, $g\in G$ \cite[Prop.~4.3]{riemLie}.
    This proves \cref{eq:metric_log} since on a neighborhood $V$ of $0\in\lie$, such that the restriction of $\exp$ to $V$ becomes a diffeomorphism, the shortest path is always the exponential one.
    We only have to prove that $d(g,h)$ is the Riemannian distance for $h=1$.
    For $n$ large enough, let $\paren{Y\up n_{1},\dots,Y\up n_{k_n}}$ be a sequence of families of generators such that $g = \rme^{Y\up n_{1}} \cdots \rme^{Y\up n_{k_n}}$ and such that $d_n \equiv \sum_j\norm{Y\up n_{j}}_\lie \to d(g,1)$.
    Without loss of generality, we set $k_n =n$.
    Then the curve $\gamma_n(t):[0,n] \to G$ with
    \begin{equation}
        t\mapsto \rme^{\tau Y\up n_{k}} \rme^{Y\up n_{k-1}} \cdots \rme^{Y\up n_{1}}
        \quad
        \text{with $(k,\tau)\in \set{1,\cdots,n}\times [0,1)$, s.t.\ $k-1+\tau =t$},
        \label{eq:w_curve}
    \end{equation}
    joins the neutral element $1$ and $g$ and has Riemannian length equal to $d_j$. Therefore the definition of $d(g,1)$ equals that of the Riemannian distance except for the fact that one allows only for piecewise geodesic curves. In particular, this shows the ``$\geq$'' part of \cref{eq:riemannian_metric}.
    For the converse, let $\gamma_n$ be a minimizing sequence of curves for \cref{eq:riemannian_metric}. Without loss of generality, we can let all $\gamma_n$ be defined on the unit interval $[0,1]$. Put $g_{n,j} = \gamma_n(j/n)$. 
    For sufficiently large $K$, there are $Y\up n_{j}= \log(g_{n,j})$ for all $j\leq n$ and $n\geq K$. 
    We now consider the curves $\tilde\gamma_n$ defined as in \cref{eq:w_curve}.
    It is clear that $L[\tilde\gamma_n]\leq L[\gamma_n]$ because we only replaced parts of the curve with geodesic, hence shorter, parts.
    Therefore, also $\tilde\gamma_n$ is a minimizing sequence, which, however, is already taken into account in the definition of $d$ in \cref{eq:metric_def}.
\end{proof}

\begin{proof}[Proof of \cref{thm:lemma_nelson}]
	\ref{it:nelson_basis_dep}: If $\{Y_j\}_j$ is another orthonormal basis, then there is an orthogonal matrix $J \in \O(m,\RR)$, where $m$ is the dimension of $\lie$, such that $Y_j = \sum_k J_{jk}X_k$. It follows that
    \begin{equation}
        \sum_j \dU(Y_j)^2 = \sum_{jkl} J_{jk}J_{jl} \,\dU(X_k)\dU(X_l) = \sum_{kl} \delta_{kl} \, \dU(X_k) \dU(X_l) =\Nelson.
    \end{equation}
    
    \ref{it:beta_ineq}: Given an $X\in\lie$, we can always find an orthonormal basis, such that $X$ is a multiple of, say, the first basis vector. Since both sides of \cref{eq:beta_ineq} are homogeneous w.r.t.\ scalar multiplication of $X$, we may assume that $X=X_1$ is the first basis vector of an orthonormal basis. This implies that $\dU(X)^2 \leq \Nelson$ which gives
    \begin{equation}
        \norm{\dU(X)\psi}^2 = \ip{\psi}{\dU(X)^2\psi} \leq \ip\psi{\Nelson\psi}.
    \end{equation}
    Since $\norm{X}_\lie=1$ in the case of a basis vector, the assertion is proved.

    \ref{it:equivalent_NL}: Let $P:\lie\to\lie$ be a linear map, such that $\ip X{PY}_\lie = \ip XY_\lie'$.
    When written in a basis, $P$ is just the Gram matrix of $\ip{\placeholder}{\placeholder}'_\lie$ expressed in the chosen basis and is necessarily positive semidefinite.
    Assume that $\{X_k\}_k$ is a $\ip\placeholder\placeholder_\lie$-orthonormal basis diagonalizing $P$, i.e\ $PX_k = p_kX_k$.
    Then $\{X_k'\}_k = \{p_k^{-1/2}X_k\}_k$ is a $\ip\placeholder\placeholder_\lie'$-orthonormal basis.
    Therefore, the $\ip\placeholder\placeholder_\lie$-Nelson Laplacian reads $\Delta =\sum_k A(X_k^2)$, and the $\ip\placeholder\placeholder_\lie'$-Nelson Laplacian is $\Delta' = \sum_k p_k^{-1} A(X_k)^2$.
    From this, it follows that $p_\mathrm{max}^{-1}\Delta\le \Delta' \le p_\mathrm{min}^{-1}\Delta$ with $p_\mathrm{min},p_\mathrm{max}$ denoting smallest and largest eigenvalues of $P$, respectively.
    Indeed, $c=p_\mathrm{min}$ and $C=p_\mathrm{max}$ are also the optimal constants for the estimates $c\ip XX_{\lie}\le\ip XX_{\lie}'\le C \ip XX_{\lie}$ for all $X\in \lie$.
\end{proof}

\begin{proof}[Proof of \cref{thm:main}]\hypertarget{prf:thm_main}{}
    Let $(U,\H)$ be a strongly continuous unitary representation of a connected Lie group whose Lie algebra $\lie$ is equipped with an $\Ad$-invariant inner product $\ip\placeholder\placeholder_\lie$.
    By left-invariance of the metric $d$, \cref{eq:main_bound} is equivalent to $\norm{U_g\psi -\psi}\leq \sqrt{\ip\psi{\Nelson\psi}}\,\\d(g,1)$ for all $\psi$ and all $g$.
    We start with the following bound, which follows from \cref{eq:beta_ineq}
    \begin{equation}
        \Big\Vert\rme^{-\rmi\dU(X)} \psi -\psi\Big\Vert = \Big\Vert\int_0^1 \rme^{-\rmi s\dU(X)} \dU(X)\psi \,ds\Big\Vert \leq \norm{\dU(X)\psi} \leq \norm{X}_\lie \sqrt{\ip{\psi}{\Nelson\psi}}.
    \end{equation}
    Assume that $Y_1,\dots, Y_n\in\lie$ are such that $g = \rme^{Y_1} \cdots \rme^{Y_n} $. Then we can write $U_g-\1$ as the telescoping sum
    \begin{align*}
        U_g-\1
            &= (\rme^{-\rmi \dU(Y_1)} -\1) + \sum_{k=2}^{n-1}  \,\rme^{-\rmi\dU(Y_1)} \cdots \rme^{-\rmi\dU(Y_{k-1})} (\rme^{-\rmi\dU(Y_{k} )} -\1).
    \end{align*}
    Applying the triangle inequality and using the bound above shows
    \begin{equation}
        \norm{U_g\psi-\psi}
        = \sum_{k=1}^n \ \norm{\rme^{-\rmi\dU(Y_{n-k+1})}\psi-\psi}
        \leq \sqrt{\ip\psi{\Nelson\psi}} \, \sum_{k=1}^n \norm{Y_k}_\lie
    \end{equation}
    We may now take the infimum over all finite collections $\set{Y_j}$ such that $g=\rme^{Y_1}\cdots \rme^{Y_n}$. This proves the claim as taking an infimum preserves the inequality.
\end{proof}

\begin{proof}[Proof of \cref{thm:channel_bound}]\hypertarget{prf:thm_channel_bound}{}
    Let $(U,\H)$ be a projective representation, continuous on a neighborhood $\UUU$ of $1\in G$, implementing $(\U,\H)$. That is for all $g\in G$, $\U_g(\rho)=U_g\rho U_g^*$.
    We will show that for $\psi\in\dom{\sqrt{\Nelson}}$ with $\norm\psi=1$ it holds that
    \begin{equation}
    	\norm{\U_g(\rho_\psi) - \U_h(\rho_\psi)}_1 \le 2\sqrt{\ip\psi{\Nelson\psi}} d(g,h),
    \end{equation}
    where $\rho_\psi = \ketbra{\psi}{\psi}$.
    The bound for the ECD norm then follows from \cite[Thm.~1]{Becker2021}, resp. \cref{lem:dim2}.
    
    First, note that for any $g,h\in G$
    \begin{align}\label{eq:channel_unitary_bound}
        \tfrac 1 2\norm{\U_{g}(\rho_\psi) - \U_h(\rho_\psi)}_1 &= \sqrt{1-|\ip{U_g\psi}{U_h\psi}|^2}\nonumber\\
        &\le \sqrt{2}\sqrt{1-|\ip{U_g\psi}{U_h\psi}|}\nonumber\\
        &= \inf_{\substack{\omega\in \mathrm{U}(1)}} \norm{U_g\psi - \omega U_h \psi}\nonumber\\
        &\le \norm{U_g\psi - U_h \psi}.
    \end{align}
    Because of left invariance, we can assume $h=1$.
    We roughly follow the idea of the proof of Theorem~\hyperlink{prf:thm_main}{\ref{thm:main}}.
    We start by showing the following bound for group elements of the form $g=\rme^X$:
    $$
        \norm{\U_{\rme^X}(\rho_\psi) - \rho_\psi}_1 \le 2 \sqrt{\ip\psi{\Nelson\psi}}\norm{X}_\lie.
    $$
    For $n\in\NN$, a telescoping sum argument shows
    $$\norm{\U_{\rme^X}(\rho_\psi) - \rho_\psi}_1 \le n\norm{\U_{\rme^{X/n}}(\rho_\psi) - \rho_\psi}_1.$$
    By choosing $n$ large enough and by homogeneity of $\norm\placeholder_\lie$, we can thus assume $\rme^X \in\UUU$.
    The generator $\dU(X)$ from \cref{lem:projgen} then satisfies
    \begin{align}\label{eq:integral_proj}
        (U_{\rme^{X}} - \1)\psi = -\rmi\int_0^1 U_{\rme^{sX}}\dU(X)\psi\d s
    \end{align}
    for $\psi\in\dom{\dU(X)}$.
    We can use \cref{eq:integral_proj} to find 
    $$\norm{U_{\rme^X}\psi - \psi} \le \int_0^1 \norm{U_{\rme^{sX}}\dU(X)\psi}\d s = \norm{\dU(X)\psi}\le \norm{X}_\lie \sqrt{\ip\psi {\Nelson\psi}}$$
    Together with \cref{eq:channel_unitary_bound}, this gives the desired bound.
    We can now proceed precisely as in the proof of Theorem~\hyperlink{prf:thm_main}{\ref{thm:main}}.
    For any family $(Y_1,\dots,Y_n)$ such that $g = \rme^{Y_1} \cdots \rme^{Y_n}$ one can use a telescoping sum to obtain
    \begin{equation}
        \norm{\U_g(\rho_\psi) - \rho_\psi}_1
        \leq \sum_{k=1}^n \, \norm{\U_{\rme^{Y_k}}(\rho_\psi)-\rho_\psi}_1.
    \end{equation}
    The right hand side is bounded by $2\sqrt{\ip\psi{\Nelson\psi}} \sum_{k=1}^n \norm{Y_k}_\lie$.
    Taking the infimum over all families of generators proves the bound
    \begin{equation}
        \norm{\U_g(\rho_\psi)-\rho_\psi}_1 \leq 2\sqrt{\ip{\psi}{\Nelson\psi}}.
    \end{equation}
\end{proof}

\begin{proof}[Proof of \cref{lem:dim2}]
        As mentioned in \cref{sec:projective}, \cref{lem:dim2} the case of $\dim(\H) \ge 3$ is the content of \cite[Thm.~1]{Becker2021}, while the $1$-dimensional case is trivial.
        This proof will deduce the result for $\dim(\H) = 2$ from the $3$-dimensional case.
        Since the proof does not depend on the specific dimensions, we show a slightly more abstract statement in hope of achieving a clearer presentation of the techniques.
        
        To do so, let $\H$ be of arbitrary dimension and consider the direct sum Hilbert space $\K = \H\oplus\CC$.
        We will extend $\U$, $\V$, and $H$ in a suitable sense that allows us to relate the corresponding ECD norm distance to that of $\U$ and $\V$.
        That $\dim(\K) = \dim(\H) + 1$ will allow us to apply \cite[Thm.~1]{Becker2021} to $\K$ (except for the trivial case $\dim(\H) = 1$).\\

        Every density operator $\omega\in\states{(\K)}$ has the form
        \begin{equation}
        	\omega\equiv\omega(\rho, \psi, p) = \begin{pmatrix}
            (1-p)\rho & \ket\psi\\
            \bra\psi & p
        \end{pmatrix}
        \end{equation}
        for some $\rho\in\states(\H)$, $\psi\in\H$ and $p\in[0,1]$.
        We will employ the notation also for operators of this form that are not necessarily density operators.
        By choosing an ONB $\ket k$ of $\H$, positivity of $\omega$ implies that
        \begin{equation}
        	\omega\supind k =
        \begin{pmatrix}
            (1-p)\bra k \rho\ket k & \bra k\psi\rangle\\
            \langle\psi\ket{k} & p
        \end{pmatrix}
        \ge 0.
        \end{equation}
        Therefore, $0\le \det(\omega\supind k) = p(1-p)\bra k \rho\ket k - \abs{\bra{k}\psi\rangle}^2$, so, by summing over $k$, we find $\norm{\psi}\le \sqrt{p(1-p)}\le \sqrt{p}$.

        We denote the unitary channel on $\states(\K)$ corresponding to $U\oplus 1$ as $\U_\K$ ($\V_\K$ is defined analogously).
        It acts as $\omega(\rho, \psi,p)\mapsto \omega(\U(\rho), U\psi, p)$.
        Let $\P$ be the pinching channel $\P:\omega(\rho, \psi, p)\mapsto \omega(\rho,0, p)$ associated to the direct sum decomposition of $\K$.
        By contractivity of quantum channels with respect to the trace norm, it then holds that
        \begin{equation}
        	\norm*{(\U_\K - \V_\K)(\omega)}_1 \ge \norm*{\P\circ(\U_\K - \V_\K)(\omega)}_1,
        \end{equation}
        where the right-hand side computes to
        \begin{equation}
        	\norm*{\P\circ(\U_\K - \V_\K)\rb*{\omega}}_1 = (1-p)\norm*{(\U-\V)(\rho)}_1.\label{eq:pinching}
        \end{equation}
        However, on the other hand, due to $\omega = \omega(\rho, 0, p) + \omega(0,\psi, 0) $,
        the triangle inequality implies
        \begin{align}
            \norm*{(\U_\K - \V_\K)(\omega)}_1 &\le (1-p)\norm*{(\U-\V)(\rho)}_1 + \norm*{\omega\rb*{0, (U-V)\psi, 0}}_1.
        \end{align}
        By exploiting the fact that $\norm*{\omega\rb*{0, (U-V)\psi, 0}}_1 = 2\norm{(U-V)\psi} \le 4\norm{\psi}$,
        we obtain
        \begin{align}\label{eq:2dimecdineq}
            (1-p) \norm*{(\U - \V)(\rho)}_1\le \norm*{(\U_\K - \V_\K)(\omega)}_1 \le (1-p) \norm*{(\U - \V)(\rho)}_1 + 4\sqrt{p}.
        \end{align}
		Here, the first inequality follows from \cref{eq:pinching}.
		
        Let $H$ be a Hamiltonian on $\H$ as in the statement of the theorem and extend it to the block diagonal Hamiltonian $H_\lambda := H \oplus \lambda$ on $\K$.
        Note,that $\Spec(H_\lambda) = \Spec(H)\cup\{\lambda\}$.
        We have that
        $\tr[\omega H_\lambda] = (1-p)\tr[\rho H] + p\lambda$,
        thus restricting to $\omega$ with energy at most $E$ implies that $\norm{\psi}^2\le p\le E/\lambda$.
        Therefore,
        \begin{equation}
        	\rb*{1-\tfrac E \lambda}\norm{\U - \V}_\diamond^{H,E} \le \norm{\U_\K - \V_\K}_\diamond^{H_\lambda, E} \le \norm{\U - \V}_\diamond^{H,E} + 4\sqrt{\tfrac E \lambda},
        \end{equation}
        so $\norm{\U_\K - \V_\K}_\diamond^{H_\lambda, E}\longrightarrow \norm{\U - \V}_\diamond^{H, E}$ as $\lambda\to\infty$.
        It also follows that for $E$ fixed and $\psi\neq 0$ or $p\neq 0$ there exists a $\lambda_*$, such that $\tr[\omega H_\lambda] > E$ for all $\lambda\ge \lambda_*$.
        Hence, the limit of $\norm{\U_\K - \V_\K}_\diamond^{H_\lambda, E}$ only depends on states $\omega(\rho,0,0)$ with $\tr[\rho H] \le  E$.
        However, for computing $\norm{\U_\K - \V_\K}_\diamond^{H_\lambda, E}$ we only need to consider pure states on $\K$ due to \cite[Thm.~1]{Becker2021}.
        But on $\omega(\rho, 0, 0)$, \eqref{eq:2dimecdineq} is an equality, so the same holds for $\norm{\U - \V}_\diamond^{H, E}$.
    \end{proof}

\begin{proof}[Proof of \cref{thm:covering}]
    Note that we may assume $h=1_G$ and $\tilde h=1_{\tilde G}$ because of left-invariance.
    Since we identify the Lie algebras of $G$ and $\tilde G$ via $\diff\pi$, it holds that $\pi(\tilde \rme^Y) = \rme^Y$.
    Therefore, if $\tilde g = \tilde \rme^{Y_1}\cdots \tilde \rme^{Y_n}$ then $g = \pi(\tilde g) = \rme^{Y_1}\cdots \rme^{Y_n}$.
    In other words, every family $Y_i$ contributing to the infimum for $\tilde d$ also contributes to $d$, so $\tilde d(\tilde g, 1_{\tilde G}) \ge d(g, 1_G)$.
    
    For $\eps>0$ we denote by $B_\eps$ the open ball of $d$-radius $\eps$ around $1_G$ in G.
    To show the opposite inequality, we first show that if $g\in B_\eps$, then the infimum in \cref{eq:metric_def} can be computed by only considering decompositions $g = \rme^{Y_1}\cdots \rme^{Y_n}$, such that the corresponding piece-wise exponential curve $\gamma = \gamma_{\rb{Y_1,\dots,Y_n}}:[0,n]\to G$ as defined in \cref{eq:w_curve} never leaves $B_\eps$. That is, we only need to consider curves with $\gamma(t)\in B_\eps$ for all $t\in[0,n]$.
    In order to show this, suppose there are $(Y_1,\dots,Y_n)$, such that the endpoint $g=\gamma(n)$ is in $B_\eps$, even though the curve is outside of $B_\eps$ at some time $t = k + \tau\in (0,n)$ with $k\in\NN$, $\tau\in[0,1)$. Then one finds 
    \begin{equation}
        \sum_{j=1}^n \norm{Y_j} > \sum_{j = 1}^k \norm{Y_j}_\lie + \tau\norm{Y_{k+1}}_\lie \ge d(\gamma(t), 1) \ge \eps > d(g,1).
    \end{equation}
    Therefore, families $(Y_1,\dots,Y_n)$, such that the curve $\gamma$ leaves $B_\eps$ at some point do not contribute to the infimum in \cref{eq:metric_def} defining $d(g,1)$ if $g \in B_\eps$.
    
    We now choose $\eps>0$ sufficiently small so that $\pi^{-1}(B_\eps)$ is a disjoint union of open sets, each diffeomorphic to $B_\eps$ via $\pi$. 
    Exactly one of them contains $1_{\covering }$, which we will denote by $V$.
    Let $\tilde g\in V$ and let $g = \pi(\tilde g) \in B_\eps$ with $g = \rme^{Y_1}\cdots \rme^{Y_n}$, such that the corresponding curve $\gamma$ lies entirely in $B_\eps$ as shown before.
    Since $\pi$ preserves exponentials and is diffeomorphic between $V$ and $B_\eps$ we obtain a piece-wise exponential curve $\tilde\gamma = \pi^{-1}\circ\gamma$, such that $\tilde\gamma(0) = 1_{\covering}$ and $\tilde\gamma(n) = \tilde g$. In particular, $\tilde\gamma$ gives a decomposition of $\tilde g$ into exponentials with the same sum of norms.
    As we have seen before, such sequences $(Y_1,\dots, Y_n)$ suffice for computing $d(g,1)$.
    We thus find that $\tilde d(\tilde g, 1) \le d(g,1)$.
\end{proof}

\section{Conclusion}\label{sec:conclusion}
In this paper, we presented a general method to obtain strong error bounds for unitary representations of any connected Lie group. Our method extends to unitary channel representations by means of the implementing projective unitary representations. Due to their generality, our bounds are directly applicable to practical calculations: By computing the Nelson Laplacian of the respective representation, strong error bounds immediately follow. We demonstrate this for six relevant examples in physics, namely spin-$j$ systems, free fermion models, displacement operators, quadratic bosonic Hamiltonians, $\mathrm{SU}(1,1)$ interferometry and spinless particles in relativistic quantum mechanics. For spin-$j$ systems, we are able to obtain bounds that are tight in first order. 
In case of free fermion models and quadratic bosonic Hamiltonians, our bounds show an improvement over the best-known bounds in the literature. Furthermore, the bounds for $\mathrm{SU}(1,1)$ interferometry and the error bounds for displacement operators in their unitary (not channel) representations as well as the result for spinless particles in relativistic quantum mechanics, even constitute novel bounds.
\\[22pt]

\paragraph{Acknowledgements.}
We thank R.\ F.\ Werner, A.\ Winter and M.\ E.\ S.\ Morales for suggestions and helpful discussions. 
We thank J.\ P.\ Solovej for suggesting the $\mathrm{SU}(1,1)$ example.

\paragraph{Funding information.}
LvL acknowledges support by the Quantum Valley Lower Saxony.
NG acknowledges financial support from the Spanish Agencia Estatal de Investigaci\'{o}n, project PID2019-107609GB-I00 and Spanish Plan de Recuperaci\'{o}n, Transformaci\'{o}n y Resiliencia financed by the European Union-NextGenerationEU.
AH was supported by the Sydney Quantum Academy.
DB acknowledges funding by the Australian Research Council (project numbers FT190100106, DP210101367, CE170100009).

\paragraph{Conflicts of interest.}
The authors declare no conflicts of interest.

\paragraph{Author contributions.}
All authors contributed equally to this work.

\begin{appendix}

\section{Nelson Laplacian for the metaplectic representation}
\label{appendix:Nelson_symplectic}

Notice that a choice of an orthonormal basis on $\mathrm{Sym}(2m,\RR)$ induces an orthonormal basis on $\mathfrak{sp}(2m,\RR)$ via the isometric isomorphism $X\mapsto \Omega X$.
We denote the standard matrix units by $E_{jk}$, i.e., $E_{jk}$ is the matrix with a one in the $j$-th row and $k$-th column and all other entries equal to zero.
Then, we define an orthonormal basis of $\mathrm{Sym}(2m,\RR)$ as
\begin{align}
    M_{k,\ell} & = ( E_{k\ell} + E_{\ell k} ) /\sqrt{2},\quad(\ell=2,\ldots,2m;k<\ell),\\
    D_{k} & =E_{kk},\quad(k=1,\ldots,2m).
\end{align}
Let us start the derivation of the Nelson Laplacian with a simple auxiliary lemma, which we will use throughout our computation.
\begin{lem}
	Let $H_k=Q_k^2+P_k^2$. Then the following identities hold
	\begin{align}
		(P_k Q_k + Q_k P_k)^{2}&=4Q_k^{2}P_k^{2}-8\rmi Q_k P_k-\1,\label{eq:pq+qp}\\
		P_k^{2}Q_k^{2} &=Q_k^{2}P_k^{2}-4\rmi Q_kP_k-2\1,\label{eq:p^2q^2}\\
		\left(\sum_{k=1}^{m}H_{k}\right)^{2}&=\left(\sum_{k=1}^{m}Q_{k}^{2}\right)^{2}+\left(\sum_{k=1}^{m}P_{k}^{2}\right)^{2}+2\sum_{k,\ell=1}^{m}Q_{k}^{2}P_{\ell}^{2}-4\rmi\sum_{k=1}^{m}Q_{k}P_{k}-2m\1.\label{eq:(sum_H_k)^2}
	\end{align}
	\begin{proof}
		All identities follow by bringing the respective quantities into a canonical order, where first all $Q_k$'s are collected and then all $P_k$'s using the canonical commutation relation $[Q_k,P_k]=\rmi\1$. 
		\\\\
		\cref{eq:pq+qp}:
		\begin{align}
			(P_k Q_k + Q_k P_k)^{2}&=P_kQ_kP_kQ_k+P_kQ_kQ_kP_k+Q_kP_kP_kQ_k+Q_kP_kQ_kP_k\nonumber\\
			&=4Q_kP_kQ_kP_k-3\rmi Q_kP_k-\rmi P_kQ_k\nonumber\\
			&=4Q_k^{2}P_k^{2}-8\rmi Q_k P_k-\1.
		\end{align}
		\cref{eq:p^2q^2}:
		\begin{align}
			P_kP_kQ_kQ_k &=Q_kP_kP_kQ_k-2\rmi P_kQ_k\nonumber\\
			&=Q_k^2P_k^2-2\rmi P_kQ_k-2\rmi Q_kP_k\nonumber\\
			&=Q_k^{2}P_k^{2}-4\rmi Q_kP_k-2\1.
		\end{align}
		\cref{eq:(sum_H_k)^2}:
		\begin{align}
			\left(\sum_{k=1}^{m}H_{k}\right)^{2}&=\sum_{k,\ell=1}^m Q_k^2Q_\ell^2+Q_k^2P_\ell^2+P_k^2Q_\ell^2+P_k^2P_\ell^2\nonumber\\
			&=\sum_{k,\ell=1}^m\left(Q_k^2Q_\ell^2+P_k^2P_\ell^2+Q_k^2P_\ell^2\right)+\sum_{k\neq \ell}^mQ_k^2P_\ell^2+\sum_{k=1}^m\left(Q_k^2P_k^2-4\rmi Q_kP_k-2\1\right)\nonumber\\
			&=\left(\sum_{k=1}^{m}Q_{k}^{2}\right)^{2}+\left(\sum_{k=1}^{m}P_{k}^{2}\right)^{2}+2\sum_{k,\ell=1}^{m}Q_{k}^{2}P_{\ell}^{2}-4\rmi\sum_{k=1}^{m}Q_{k}P_{k}-2m\1,
		\end{align}
		where the second step follows from \cref{eq:p^2q^2}.
	\end{proof}
\end{lem}
\noindent
Let us sum up the squares of the Hamiltonians corresponding to the non-diagonal symmetric matrices $M_{k,\ell}$ first. That is, we compute $\Nelson_M=\sum_{k,\ell}\dU(M_{k,\ell})^2$. For this purpose, define $H_{k}=(Q_{k}^{2}+P_{k}^{2})$ as before. Then
\begin{align}
\Nelson_{M} & =\frac{1}{4}\left(\sum_{k\le\ell}^{m}\frac{\left(Q_{k}P_{\ell}+P_{\ell}Q_{k}\right)^{2}}{2}+\sum_{k<\ell}^{m}\frac{\left(2Q_{k}Q_{\ell}\right)^{2}}{2}+\sum_{k<\ell}^{m}\frac{\left(2P_{k}P_{\ell}\right)^{2}}{2}\right)\nonumber\\
 & =\frac{1}{4}\left(\sum_{k<\ell}^{m}2Q_{k}^{2}P_{\ell}^{2}+\sum_{k=1}^{m}\frac{4Q_{k}^{2}P_{k}^{2}-8iQ_{k}P_{k}-\1}{2}+\sum_{k<\ell}^{m}2Q_{k}^{2}Q_{\ell}^{2}+\sum_{k<\ell}^{m}2P_{k}^{2}P_{\ell}^{2}\right)\nonumber\\
 & =\frac{1}{4}\Bigg(2\sum_{k,\ell=1}^{m}Q_{k}^{2}P_{\ell}^{2}-4\rmi\sum_{k=1}^{m}Q_{k}P_{k}-\frac{m}{2}\1+\left(\sum_{k=1}^{m}Q_{k}^{2}\right)^{2} + \dots \nonumber\\
 &\qquad \ldots +\left(\sum_{k=1}^{m}P_{k}^{2}\right)^{2}-\sum_{k=1}^{m}Q_{k}^{4}-\sum_{k=1}^{m}P_{k}^{4}\Bigg)\nonumber\\
 & = \frac{1}{4}\left(\left(\sum_{k=1}^{m}H_{k}\right)^{2}-\sum_{k=1}^{m}Q_{k}^{4}-\sum_{k=1}^{m}P_{k}^{4}+\frac{{3m}}{2}\mathds{{1}}\right),
\end{align}
where we used \cref{eq:pq+qp} in the first step and \cref{eq:(sum_H_k)^2} in the last step. For the diagonal matrices $D_k$, we have for $\Nelson_D=\sum_k\dU(D_k)^2$
\begin{equation}
	\Nelson_D=\frac{1}{4}\left(\sum_{k=1}^m Q_k^4+\sum_{k=1}^m P_k^4\right).
\end{equation}
Thus, the Nelson Laplacian $\Nelson=\Nelson_M+\Nelson_D$ becomes
\begin{equation}
	\Nelson=\frac{1}{4}\left(\sum_{k=1}^m H_k\right)^2+\frac{3m}{8}\1.
\end{equation}

\section{Nelson Laplacian for the FLO representation}\label{appendix:Nelson_SO}

Recall that we equipped $\mathfrak{so}(2m,\RR) = \set{B \in \RR^{2m\times2m} \given B^\top =-B}$ with the Frobenius inner product.
We choose the following orthonormal basis
\begin{align}
	B_{jk}&=(E_{jk}-E_{kj})/\sqrt{2}, \quad j=1,\dots, 2m,\quad k<j,
	\label{eq:so_basis}
\end{align}
where the $E_{jk}$ are defined as in \cref{appendix:Nelson_symplectic}. Inserting this back to \cref{eq:Hamiltonian_free_fermion} gives
\begin{align}
	\dU(B_{jk})&=\frac{\rmi}{4\sqrt{2}}[c_j,c_k], \quad j=1,\dots, 2m,\quad k<j.\label{eq:free_fermion_basis}
\end{align}
\cref{eq:free_fermion_basis} can be further simplified by using the fact that $[c_j,c_k]=2(\delta_{jk}\1+c_kc_j)$
\begin{equation}
	\dU(B_{jk})=\frac{\rmi}{2\sqrt{2}} c_kc_j, \quad j=1,\dots, 2m,\quad k<j.
\end{equation}
From \cref{eq:Clifford}, we infer for $j\neq k$ that $c_jc_k=-c_kc_j$ and it follows
\begin{equation}
	\dU(B_{jk})^2=\frac{1}{8}c_kc_kc_jc_j=\frac{1}{8}\1,
\end{equation}
where we have used $c_jc_j=\1$, $\forall j$. Since $\mathrm{dim}(\mathfrak{so}(2m))=m(2m-1)$, the Nelson Laplacian reads
\begin{align}
	\Nelson=\frac{m(2m-1)}{8}\,\1.
\end{align}

\end{appendix}

\bibliography{LieErrorBounds_JPhysA.bib}

\nolinenumbers

\end{document}